\documentclass[useAMS,usenatbib]{mn2e}
\usepackage{graphicx,fleqn}
\DeclareGraphicsExtensions{.eps,.ps,.eps.gz,.ps.gz,.eps.Z}
\def \apj{ApJ}
\def \apjl{ApJL}

\def \aap{A\&A}

\def \mnras{MNRAS}

\input psfig.sty

\title[Jet breaks at the end of ...]{Jet breaks at the end of the slow
decline phase of \textit{Swift} GRB lightcurves }

\author[M. De Pasquale et al.]{M. De
Pasquale$^{1}$\footnote{Send reprint requests to:
mdp@mssl.ucl.ac.uk}, P. Evans$^{2}$, S. Oates$^{1}$, M. Page$^{1}$,
S. Zane$^{1}$, P. Schady$^{1}$, \newauthor A. Breeveld$^{1}$,   S. Holland$^{3}$, P. Kuin$^{1}$, M. Still$^{1}$, P. Roming$^{4}$, P. Ward$^{1}$  \\
$^{1}$ Mullard Space Science Laboratory, University College London,
Holmbury St. Mary, Dorking Surrey, RH5 6NT, UK;  \\
$^{2}$Department of Physics and Astronomy, University of Leicester,
Leicester, LE1 7RH, UK; \\
$^{3}$Goddard Space Flight Center - Goddard Space Flight Center, Greenbelt, MD 20771, USA. \\
$^{4}$Department of Astronomy and Astrophysics, Pennsylvania State
University, 525 Davey Lab, University Park, PA 16802, USA
 }

\begin{document}

\date{Accepted...Received...}

\maketitle

\label{firstpage}

\begin{abstract}

The Swift mission has discovered an intriguing feature of Gamma-Ray
Burst (GRBs) afterglows, a phase of shallow decline of the flux in
the X-ray and optical lightcurves. This behaviour is typically
attributed to energy injection into the burst ejecta. At some point
this phase ends, resulting in a break in the lightcurve, which is
commonly interpreted as the cessation of the energy injection. In a
few cases, however, while breaks in the X-ray lightcurve are
observed, optical emission continues its slow flux decline. This
behaviour suggests a more complex scenario. In this paper, we
present a model that invokes a double component outflow, in which
narrowly collimated ejecta are responsible for the X-ray emission
while a broad outflow is responsible for the optical emission. The
narrow component can produce a jet break in the X-ray lightcurve at
relatively early times, while the optical emission does not break
due to its lower degree of collimation. In our model both components
are subject to energy injection for the whole duration of the
follow-up observations. We apply this model to GRBs with chromatic
breaks, and we show how it might change the interpretation of  the
GRBs canonical lightcurve. We also study our model from a
theoretical point of view, investigating the possible configurations
of frequencies and the values of GRB physical parameters allowed in
our model.

\end{abstract}


\section{Introduction}
\label{intro}

Since its launch, the \textit{Swift} mission (Gehrels et al. 2004)
has allowed us to observe the emission from Gamma-Ray Burst (GRB)
afterglows in the X-ray and UV/Optical from as early as $\sim$1
minute after the burst trigger by means of the X-ray Telescope (XRT,
Burrows et al. 2004) and UV/Optical Telescope (UVOT, Roming et al.
2005). This unprecedented response time has allowed us to unveil the
early behaviour of GRB afterglow lightcurves, which turn out to be
more complex than expected. Typically, at the end of the prompt
emission the X-ray flux $F$ exhibits a rapid decay. This can be
modeled with a powerlaw $F\sim t^{-\alpha{_1}}$ with slope
$\alpha_{1} \sim 3 - 5$ (Tagliaferri et al. 2005). This phase, which
usually lasts hundreds of seconds, is widely interpreted as the tail
of the prompt emission (Kumar \& Panaitescu 2000; for a review, see
Zhang et al. 2006). After that, the X-ray flux decays in a much
shallower way, forming a ``plateau'' with a slope $\alpha_{2} \sim
0.1 - 0.8$. The spectrum in this phase can be different from that
observed during the fast decay, which indicates a different physical
origin. The duration of the slow decline is a few thousands of
seconds (O'Brien et al. 2006, Willingale et al. 2007). After this
time, a break occurs and the lightcurve becomes steeper, with a
powerlaw slope of $\alpha _{3}\sim1.3$. Indeed, this latter phase
was studied well prior to the launch of \textit{Swift} (e.g. De
Pasquale et al. 2006; Gendre et al. 2006) and it is understood to be
emission from synchrotron radiation, resulting from a shock produced
by the expansion of the burst ejecta into the circumburst medium
(Meszaros \& Rees 1997). Occasionally, a further break may occur a
few days after the trigger, leading to a segment with decay slope of
$\alpha_{4}\sim2$ . This steep decay can be interpreted as the
signature of collimated outflow (Sari et al. 1999). Overall, this
evolution of the X-ray flux  is now referred as the ``canonical''
X-ray lightcurve (Nousek et al. 2006). In the optical band, the flux
decays with a similar range of slopes to those of the X-ray, with
the exception of the
initial fast decay phase, which is usually absent (Oates et al. in preparation).\\

 The slow decay is probably the most perplexing among the novel
aspects discovered by \textit{Swift}, and several models have been
proposed to explain it (see e.g. Zhang 2007 for a complete review).
These models in general fall into three main classes: i) energy
injection into the burst ejecta, either in the form of Poynting flux
or late time shells of jecta; ii) a non uniform angular energy
distribution in the jet or a jet seen off-axis, so that a fraction
of the early afterglow emission is not fully beamed towards the
observer; iii) a change of the microphysical parameters that leads
to an increase in the conversion efficiency of the ejecta energy to
radiation.
% In the majority of GRBs, achromatic breaks are observed,
%both in the X-ray and in the optical bands, between the slow decay
%phase (with index $\alpha_2$) and the normal decay phase (with index
%$\alpha_3$).

 Puzzlingly, in a few \textit{Swift} GRBs the slow decline phase ends
with a ``chromatic break'' (Panaitescu et al. 2006a; see also
Melandri et al. 2008): i.e. a transition from the shallow to the
normal decay appears in the X-ray band but is absent in the optical
band, where the flux continues to decline at a slow rate. This
feature is very hard to explain with any model that predicts a
single origin for the X-ray and optical emission. In the attempt to
solve this problem, Ghisellini et al. (2007) suggested a model in
which the optical emission is caused by the interaction between the
ejecta and the circumburst medium, while the X-ray radiation is
produced by internal shocks occurring in collimated shells emitted
by the GRB central engine at relatively late times. If the Lorentz
factor $\Gamma$ of these shells decreases with time, a ``jet-like''
break will be detected (in the X-ray band only) at the time in which
$\Gamma^{-1} = \theta$, where $\theta$ is the opening angle of
ejecta.
%This model, however, is not easily reconciled with the
%observed spectral changes between the fast and slow decay phases.
 An alternative scenario, proposed by Genet et al. (2007) and Uhm \&
Beloborodov (2007), assumes that both the X-ray and optical emission
is due to reverse shocks crossing the shells.
% In this scenario, if
%the Lorentz factor of late shells decreases down to $\Gamma \sim$ a
%few and only a small fraction ($<5\%$) of the kinetic energy is
%transferred to the electron population, the model predicts
%lightcurves similar to those observed.
 However, this model requires that the external shock emission is
basically turned off. This may need conditions difficult to meet.
Other authors argue that the jet breaks are actually hidden in the
optical lightcurves (Curran et al. 2007) and/or less evident than
expected (Panaitescu et al., 2007a, Liang et al. 2007). In
Panaitescu (2007b), the author proposes a complex scenario, in which
the plateau, the flares and the chromatic breaks seen in the X-ray
lightcurve are caused by scattering of the forward-shock synchrotron
emission by a relativistic outflow, located behind the leading
blast-wave. Efforts have also been made to reconcile the chromatic
breaks with the scenario of an unique outflow (Panaitescu et al.
2006a), hypothesizing an evolution of the microphysical parameters,
including the fractions of blast wave energy given to electrons and
to the magnetic field. However, as the authors themselves pointed
out, the required evolution is assumed ``ad hoc'', and still lacks a
self-consistent physical explanation.

Recently, Oates et al. (2007) have investigated the case of
\textit{Swift} GRB050802, one of the bursts in the dataset of
Panaitescu (2006a), which shows a very evident chromatic break. They
found that the observed late SED cannot be reproduced by models
based on single component outflow, and proposed a model based on two
outflows: a narrow one responsible for the X-ray emission, and a
wider one that powers the optical emission. Both outflows receive
continuous energy by means of shells emitted at late times or in the
form of Poynting flux. The break in the X-ray lightcurve, in this
scenario, is interpreted as a jet break, and there is no
discontinuation of energy injection. The ``normal'' decay phase is
then a post jet break phase with a slope less steep than usual
because of the energy injection. The fact that the optical
lightcurve does not show a break within the time of the follow-up
observations is naturally explained by the lower degree of
collimation of the outflow responsible for it. In this paper, we
conduct a detailed analysis of a sample of other GRBs that are
reported to have chromatic breaks, showing that this model can
potentially interpret the observed behaviour. We also discuss how
this scenario may change our interpretation of the canonical
lightcurve of GRBs and the deep implications that this change of
perspective may have on our understanding of GRB physics. This paper
is organized as follows. In \S~\ref{data_analysis} we introduce the
dataset and the data analysis,  while in \S~\ref{res} we present the
application of the model to the GRB sample. Discussion and
conclusions follow in \S~\ref{disc} and \S~\ref{conc}, respectively.

\section{Data Reduction and Analysis}
\label{data_analysis}

In this work, we reexamine all \textit{Swift} GRBs with chromatic
breaks contained in the sample of Panaitescu et al. (2006a), namely
GRB050319, GRB050401, GRB050607, GRB050713, GRB050802, GRB050922c,
in the light of the results found by Oates et al. (2007) on
GRB050802. We also include in our analysis \textit{Swift} GRB060605,
which is another example of a burst with a chromatic break and good
quality data.

As we will discuss later on, while the X-ray analysis alone can
indicate that our model is compatible with the observations, the
presence of a second outflow can be robustly confirmed only by a
joint analysis of the X-ray and optical data. In this respect, we
note that two bursts in the Panaitescu's dataset, GRB050607 and
GRB050713,  have poorly sampled optical  data, while for a further
one, GRB050401, no UVOT data are available because of the presence
of a bright star in the field of view. For these events, we will
only consider the X-ray emission, to show that our scenario is fully
consistent with the observations.

Once a GRB has been detected by the BAT, Swift immediately slews,
allowing the XRT and UVOT to provide prompt simultaneous multi-band
data. In the following, we describe how XRT and UVOT data are
reduced and analysed.

\subsection{XRT data reduction}

To determine the X-ray properties of the GRBs, we first re-ran the
processing pipeline version 2.72 of the Swift software. We generated
light curves using the software of Evans et al. (2007) which
supplies the \textit{Swift} XRT light curve
repository\footnote{http://www.swift.ac.uk/xrt\_curves}, and
modelled them with a sequence of connected powerlaw decays, using
$\chi^2$ minimization. In this way we identified the segments of the
lightcurves corresponding to the lightcurve segments of the
canonical X-ray lightcurve. We then extracted spectra and effective
area files (ARFs) for the plateau and post-plateau phases. Where the
source was piled up, we fitted the source PSF profile with
\emph{Swift}'s known PSF (Moretti et al. 2006) to determine the
radius within which pile-up is important, and used an annular
extraction region so that data from the piled-up part of the PSF was
excluded. If the source was not piled up, we used a circular
extraction region of 20 pixel radius (or smaller for faint sources,
to maximise the signal-to-noise). In some cases, a single
light-curve segment could cover several decades of count-rate, with
pile up only being a problem at the start of the segment. In these
cases we extracted two event lists, using  an annular source region
when pile-up occurred and a circular one at all other times, and
created separate ARF files for the two extraction regions. The event
lists were then combined using {\sc xselect} and a single spectrum
was generated from the extracted events; the ARFs were merged using
the {\sc addarf} tool, and weighting the component ARFs by source
count-rate. Background spectra were always extracted from an annulus
centred on the source; these annuli were searched for sources, and
any found were excluded from the extraction region. Where a light
curve segment spanned multiple \emph{Swift\/} observations, separate
event lists and ARFs had to be generated for each observation; these
were also combined as just described. Where a spectrum corresponding
to a specific time was required to produce a combined UVOT+XRT
spectral energy distribution, we first determined the count rate $C$
at the epoch of interest from the best fit parameters of the light
curve, then we modified the exposure time in the spectral file so
that the
resulting count rate was equal to $C$.\\

\subsection{UVOT data reduction}

 UVOT observes the GRB field through a number of pre-planned exposures.
The automatic target(AT) sequence begins with a short settling
exposure followed by either one or two finding charts. UVOT performs
observations either in event mode, where the position and arrival
time of each photon is recorded; or, in image mode, where an image
is accumulated over a fixed period of time. The GRB is expected to
vary over the shortest timescales during the first few hundred
seconds after the trigger; therefore, the settling exposure and
finding charts are observed in event mode. The rest of the AT
sequence contains a series of exposures, in the 7 filters, lasting
from as little as 10s through to a few thousand seconds. These are
observed through a combination of event (until $\sim$850s after the
trigger) and image mode observations.

 The aspect and astrometry for each photon, in the case of the event
data, was refined following the method of Oates et al. (in prep.).
The images were processed by the pipeline at the Swift Science Data
Center (SDC). Any images not aspect corrected during the pipeline
processing were corrected using bespoke aspect correction software.
To produce lightcurves, the source counts were extracted in an
aperture which was sized according to the count rate. For count
rates higher than 0.5 counts per second, a 5$\arcsec$ radius circle
was used, and for count rates lower than 0.5 counts per second the
source count rates were obtained using a 3$\arcsec$ radius circle,
and were then corrected to 5$\arcsec$ using the PSFs recorded in the
calibration files (Poole et al. 2007). The background count rates
were determined using a circle of radius 20$\arcsec$, positioned
over a blank area of sky near the source position. The software used
to extract the count rates can be found in the software release,
Headas 6.3.2 and version 20071106(UVOT) of the calibration files. In
order to produce a single optical light curve for each GRB in the
sample, the lightcurves in each UVOT filter were renormalised to
that in the V filter. The normalisations were determined by
performing a simultaneous power law fit, in which the lightcurves in
the different filters have the same slope but were allowed different
normalisations, in periods in which the lightcurve can be described
as a powerlaw decay. The count rates from each filter were then
binned by taking the weighted average in
time bins of $\delta$T/T\,=\,0.2.\\

 In order to understand the properties of GRBs of our sample, we built
the Spectral Energy Distributions (SEDs) at two epochs, before and
after the end of the plateau in the X-ray lightcurve. As for the
optical, we used the best fit normalisation for each filter to get
the corresponding count rate at the epoch of interest, by using the
best fit decay index. The uvottools ``uvot2pha'' and ``ftedit'' were
used to create the spectral files and convert the count rate to the
value determined in the lightcurves fitting described above.

\subsection{Spectral modelling}

All spectra were fitted in {\sc XSPEC 12.3}.
 The X-ray spectra were binned to contain a minimum of 15 counts per bin
(20 counts for the brightest spectra), and we used the version 10
response files (Godet et al. 2008). Some of the plateau-phase data
comprised both Windowed Timing (WT) and Photon Counting (PC) data,
in which case the two modes' spectra were fitted together with the
same model, but a (free) constant factor applied to the
normalisation.

 Theoretical predictions and observational findings indicate that the
spectral shape of a GRB afterglow is typically either an unbroken or
a broken powerlaw throughout the X-ray and optical bands. The break
frequency is the synchrotron cooling break, $\nu_{C}$, in which case
the difference in the spectral slopes of the broken powerlaw is 0.5.
Therefore, we jointly fitted the optical and X-ray SED with two
models. One model consisted of unbroken powerlaw, two absorbers and
two dust models (zdust in \textit{Xspec}). The column density of one
of the two absorbers was fixed to the Galactic value at the
coordinates of the GRB, given by Kalberla et al.~(2005), while the
value of reddening in one of the zdust model was frozen to the value
derived from the absorber value, according to relation between
$E_{B-V}$ and the hydrogen column density (Bohlin et al. 1978). The
redshift of the other absorber and zdust component was fixed to the
corresponding burst redshift.\footnote{In this regard, all the
bursts for which we built the optical and X-ray SEDs have their
redshift known by spectroscopy}. The second model was different only
in substituting the the powerlaw with with a broken powerlaw, with
the second spectral slope bound to be higher than the first by 0.5.
 In the process of spectral analysis, we tried the Galactic, Large
Magellanic Cloud and Small Magellanic cloud extinction laws (Gal,
LMC and SMC henceforth). However, since in all cases (apart from
GRB050802, see below) it has been impossible to disentangle among
these three extinction laws, in the following we report results
obtained adopting the SMC extinction law, which provides acceptable
results in the fits of the extinction laws of the GRB host galaxies
(Stratta et al. 2004, Schady et al. 2007). For spectral modelling of
those bursts which only have X-ray data, the model was reduced to a
single powerlaw and the two photoelectric absorbers.

In the following sections of this paper, we use the convention
$F\sim t^{-\alpha} \nu^{-\beta}$ and errors are indicated at $68\%$
confidence level (c.l.). The subscripts ``O'' and ``X'' refer to
optical and X-ray respectively. We will add the labels ``1'', ``2'',
etc to attribute the decay and spectral slope to the relative
portion of the canonical X-ray lightcurve. The segments of the X-ray
canonical lightcurve which will thus be $\alpha_{X,1}$,
$\alpha_{X,2}$, $\alpha_{X,3}$. The time when the breaks in the
X-ray lightcurve occur will be indicated as $t_{X,1}$ and $t_{X,2}$.
$\alpha_{X,2}$ will always be the decay slope of the slow decaying
segment. $\alpha_O$ will be the slope of the optical lightcurve. If
any break is detected in this band, we will define $\alpha_{O,1}$
and $\alpha_{O,2}$ as the pre and post break slope, respectively.
The labels $\beta_{X,1}$, $\beta_{X,2}$ and $\beta_{X,3}$ will
indicate the spectral energy slopes of the X-ray data only.  As for
the analysis of the SED, in the case of a fit with single powerlaw,
$\beta_{OX}$ is the energy index of the spectrum. In the case of a
fit with broken powerlaw, we shall use two energy indeces, which
will be referred to as $\beta_{OX}$ and $\beta_{OX}+0.5$ (we remind
that the difference between the two indeces is fixed to be 0.5).
Additional ``E'' and ``L'' labels indicate if the fit was performed
before or after the break in the X-ray. \\The results of the
temporal analysis of the GRBs are given in Table \ref{tab1}, and
those of the spectral analysis are reported in Tables \ref{tab2},
\ref{tab1a} and \ref{tab5}. The formulae we shall be using are
recollected in Tab. \ref{Table1}.

\section{Results of GRB data analysis.}
\label{res}

\subsection{GRB050319}

\begin{figure*}
 \includegraphics[width=1.0\textwidth=0.8]{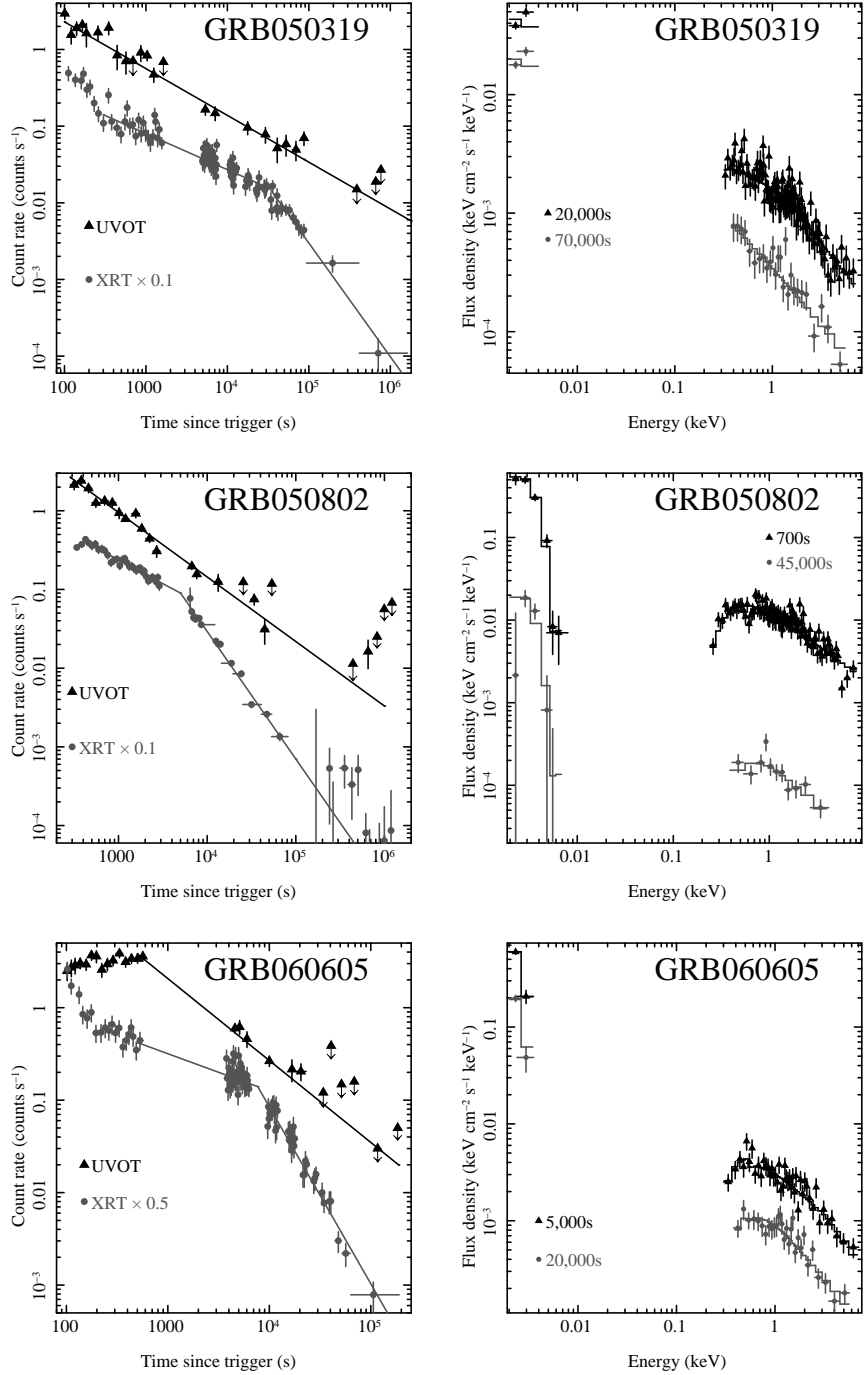}
 \caption{Lightcurves and SEDs of GRB050319, GRB050802, GRB060605.
UVOT lightcurves are fitted by simple powerlaws, while XRT
lightcurves and SEDs are fitted by broken powerlaws}\label{fig4}
\end{figure*}

The X-ray lightcurve of GRB050319 (Fig \ref{fig4}, top panel) shows
the typical canonical behaviour, and can be adequately fitted by a
double broken powerlaw model, which yields $\chi=113.3$ for 113
d.o.f.. The best fit parameters are decay indices of $\alpha_{X,1} =
1.45 ^{+0.10} _{-0.11}$, $\alpha_{X,2}=0.48 \pm 0.03$, $\alpha_{X,3}
= 1.41 ^{+0.08} _{-0.07}$ and break times among these segments of
$t_{X,1} = 300 ^{+60} _{-30}$ s and $t_{X,2} = 29.9 ^{+2.6} _{-2.8}$
ks. The early, relatively steep decay is likely the tail of the
prompt emission, a mechanism that does not involve the forward
shock; we will therefore ignore this part of the emission hereafter.
The initial flat decay phase, between $t_{X,1}$ and $t_{X,2}$, has a
spectrum with a powerlaw index $\beta_{X,2} = 1.00\pm0.03$. After
$t_{X,2}$, the X-ray spectrum shows marginal indication of
softening, since the best fit index is $\beta_{X,3} = 1.12\pm 0.07$.

The fit of the optical lightcurve (Fig.\ref{fig4}, top panel) with a
single powerlaw provides a marginally acceptable fit, yielding
$\chi^{2}=54.6$ with 25 d.o.f. The best fit decay index is
$\alpha_{O} = 0.62 \pm 0.02$. A fit with a broken powerlaw is
slightly better, yielding $\chi^{2}=43.2$ for 23 d.o.f. The F-test
indicates that the probability of chance improvement is very
marginal, $6.5\%$. As for the broken powerlaw model, the values of
the best fit parameters, other than the first slope, are not well
constrained, if we leave all of them free to vary. We then fixed the
value of the second slope, forcing it to differ from the first decay
slope as much $\alpha_{X,2}$ differs from $\alpha_{X,3}$. We thus
obtained $\alpha_{O,1} = 0.58 \pm {0.03}$, $\alpha_{O,2} = 1.52$ for
the two decay indices, and a break time $t_{O} = 164.4^{+104.7}
_{-79.5}$~ks. To find a strong upper limit on $t_{O}$, we varied its
value while fitting the other parameters, until we obtained
$\Delta\chi^{2}$ of 9. We found that we have $t_{O} > 51.5$~ks at 3
sigma confidence level. Therefore, we note that a break in the
optical band, if any, takes place much later than the break in the
X-ray. Our result are consistent with those of Panaitescu et al.
2006, in which the authors do not find any steepening of the optical
band emission up to $\sim400$ks after the trigger. All these
findings indicate that GRB050319 has got a genuine chromatic break
in the X-ray band only at about 30ks after the trigger.
% With the same procedure described
%above, we found a 90\% Confidence Level (C.L.) upper limit of
%$\alpha_{O,2}>1.1$.
% and changing the  We do not
%report the errors in the values of the second slope (after $t_{O}$),
%since this parameter is not well constrained. However, we froze the
%other parameters and varied the value of $\alpha_{O,2}$ until we
%obtained a variation of the $\chi^{2}$ of 2.7. By doing this,
%The F-test
%indicates that the probability of a improvement by chance is 4.5\%.
%Therefore, while we cannot exclude the possibility that optical
%lightcurve of GRB050319 is a simple powerlaw, we do however favour
%the fit with a broken powerlaw. The best fit decay slope for the simple powerlaw model is.
% However, we can
%show that the break in the optical, if present, occurred much later
%than the break in the X-ray. In fact, if the second index is forced
%to be the same as the third segment of the X-ray lightcurve, we can
%vary the break time until we obtain a variation of
%$\Delta\chi^{2}=2.71$ and $\Delta\chi^{2}=9$ with respect to the
%best fit with a broken powerlaw. By doing this, we found a lower
%limit of $t_{O,b}= 65$~ks and 43.8~ks respectively, significantly
%later than the second break in the X-ray. We therefore conclude that
%GRB050319 has a genuine chromatic break in the X-ray only, at 29.91
%ks.

 The SEDs of GRB050319 were built at 20ks and 70ks after the trigger
(Fig.~\ref{fig4}, top panel); results of the fit are shown in
Table~\ref{tab2}. For both SEDs, the fit with a cooling break in the
spectrum yields a better $\chi^{2}$ than the fit with a single
powerlaw, which is nevertheless still acceptable. In the following
we will discuss both the cases of unbroken and broken powerlaws.

 Let us first consider a scenario in which the X-ray and optical
bands lie on the same spectral segment at 20ks, below the cooling
frequency. This corresponds to the spectral fit with a single
powerlaw of slope $\beta_{OX,E} = 0.84\pm0.05$. In this scenario,
one should expect that the fluxes of both bands decay with the same
slope. We find that the X-ray slope observed at early times,
$\alpha_{X,2}=0.48 \pm 0.03$, is consistent within $\sim 2.4\sigma$
with $\alpha_{O,1}=0.58\pm0.03$. The average decay index of X-ray
and optical is $\overline{\alpha}=0.53\pm0.02$. Such a shallow
optical decay requires that energy injection takes place. The value
of the energy injection parameter $q$ is linked to the values of the
spectral and decay indices ($\beta$ and $\alpha$) through the
expression collected in Tab.~\ref{Table1} (Zhang et al.~2006,
Panaitescu et al. 2006b). In the case at hand, we have $q = 0.50 \pm
0.06$
% $q (\overline{\alpha} -\beta_{OX,E} +1)(1+1/2
%\beta_{OX,E}) = 0.51 \pm 0.06$,
in the standard hypothesis of a constant density environment (ISM).
The break in the X-ray lightcurve at 30~ks is generally interpreted
as the cessation of energy injection. However, if this were the
right scenario, the optical emission decay slope would
simultaneously increase up to $\alpha=3\beta_{OX,E} / 2 = 1.26 \pm
0.08$, similar to the X-ray decay slope. This prediction is not
consistent with our analysis. Alternatively, if the 30 ks break in
the X-ray band were due to the transit of the cooling frequency
below the X-ray band and not to the end of energy injection, the
expected post-break decay index would be (Tab.\ref{Table1}) $\alpha
= 0.95 \pm 0.08$
% $\alpha = (q-2)/2 + \frac{2+q}{2} (\beta_{OX,1}+0.5) = 0.62 \pm
%0.08$,
whereas the observed value is $\alpha_{X,3} = 1.41 ^{+0.08}
_{-0.07}$.
% This amount of energy injection
%cannot be reconciled with the late steep X-ray decay: even if the
%cooling frequency transited below the X-ray at late times, if $q =
%0.51 \pm 0.06 $
 Another possibility would be that the cooling frequency is
already between the optical and the X-ray bands at the time of the
first SED. This corresponds to the broken powerlaw fits, where we
find a low energy spectral slope $\beta_{OX,E}=0.49\pm0.05$ at 20~ks
and $\beta_{OX,L}=0.58^{+0.19} _{-0.12}$ at 70ks. The corresponding
high energy spectral slopes are set to be higher by 0.5. If the
cooling break is between the two bands, the only scenario that can
explain why the X-ray flux decays slower than the optical, before
the break at 30ks, is one in which the density profile of the
circumburst medium is typical of a wind ejected by a massive star
(with density decreasing as $r^{-\delta}$ where $r$ is the distance
from the centre of the explosion and $\delta \sim 2$; see Chevalier
\& Li 2000). However, even this scenario cannot explain the decay
slopes of X-ray flux and optical emissions after the 30ks break. In
fact, the conventional interpretation of the canonical X-ray
lightcurve is that after the 30~ks break the ejecta do not undergo
any further increase of their kinetic energy. Without energy
injection, in a wind environment, the decay slope above the cooling
frequency would be less steep than that of the optical by $0.25$,
which is obviously not in agreement with our observations. For
example, the optical slope we would expect is $\alpha
=\frac{3}{2}\beta +1/2$, where $\beta$ is the spectral slope in this
band. Taking $\beta$ as the weighted average of the low energy
spectral slopes, the optical decay should be
$\alpha_{O}=1.25\pm0.08$, and the X-ray decay should be $\alpha_X=
\alpha_O - 0.25 = 1.00\pm0.08$, which is evidently in contrast with
our findings.
In summary, this shows that the steep late X-ray decay is not
explained if we assume that X-ray and optical are originated by
the same component.\\
We can now demonstrate that the late X-ray break can be easily
explained as a jet break, under the assumptions that the outflow
responsible for the X-ray is different from that producing the
optical emission, and the energy injection rate does not change till
the end of the observations. Here and in the following, we will only
consider the simple case of side-spreading jet and constant density
medium with the addition of energy injection (Panaitescu et al.
2006b, but see \S~\ref{disc} for a discussion). In such a model, the
energy is assumed to increase as a simple powerlaw, $E\propto
t^{(1-q)}$, and the energy injection parameter $q$ does not change
with time. To compute the value of $q$, we need the decay and
spectral slopes, as in previous cases. In the X-ray, decay index is
$\alpha_{X,2}=0.48\pm0.03$, while for the spectral index we can take
the weighted average of the energy index found by the X-ray data
analysis throughout the whole lightcurve, $\overline{\beta}_{X} =
1.02\pm0.03$. With these values of parameters and in the case of the
X-ray band above $\nu_C$, we derive (Tab.~\ref{Table1}) $q = 0.46
\pm 0.06$. If there were not such energy injection, the decay slope
after the jet break would become $\alpha = 2\beta = 2.04\pm0.06$;
but the addition of energy into the blastwave flattens the slope,
leaving the flux decaying with $\alpha=1.31\pm0.10$, a value which
is within $1\sigma$ from the observed one,
$\alpha_{X,3}=1.41^{+0.08} _{-0.07}$.
%If the X-ray band is above the
%cooling frequency $\nu_{C}$ for the whole duration of the
%observation,
% $q$ is linked to the values of the spectral and decay
%indices ($\beta$ and $\alpha$) in the slow decline phase though the
%expression (Zhang et al.~2006) $ q = 2 (\alpha -\beta +1)/(1+\beta)
%$
%\begin{equation}
% q= 2 (\alpha -\beta +1)/(1+\beta) \, .
%\label{q1}
%\end{equation}
%The decay slope is $\alpha_{X,2}=0.48\pm0.03$, while for the
%spectral index we can take the weighted average of the energy index
%found by the X-ray data analysis throughout the different segments
%of the lightcurve, $\overline{\beta}_{X} = 1.02\pm0.03$. This gives
%$q = 0.46 \pm 0.07$. Now, in absence of energy injection the
%expected decay slope after the jet break would be $\alpha=2
%\overline{\beta}_{X}= 2.04\pm0.06$
%\begin{equation}
%\alpha=2\beta \, ,
%\label{al1}
%\end{equation}
%which gives  $\alpha= 2.04\pm0.06$.

%by a factor (Panaitescu
%et al.,~2006)
%\begin{equation}
%\Delta \alpha =0.67(1-q)(1+\beta) \label{dal1}
%\end{equation}
%thus leaving the flux decaying with $\alpha=1.31\pm0.13$, a value
%which is within $1\sigma$ from the observed one.

 In order to compute the size of the beaming angle, $\theta$, of the narrow
outflow, we use the expression (Frail et al. 2001):

\begin{equation}\label{9}
\theta = 0.093 \left (\frac{t_{j,d}}{1+z} \right )^{3/8}
{E_{k,52}}^{-1/8} \left (\frac{n}{0.1} \right )^{1/8} \, {\rm rad}\,
,
\end{equation}
where $t_{j,d}$ is the jet break time in days, $E_{k,52}$ is the
isotropic kinetic energy of the outflow, and $n$ is the density of
the environment in protons per cubic centimetre.
%For values of the
%parameters, $\eta=0.05$, $n=0.01$~cm$^{-3}$ (Frail et al. 2001),
%this gives

 As we will discuss later on (see \S~4), in order for our model to
hold the kinetic energy in the outflow responsible for the X-ray
emission should be of order $\sim10\%$ of the whole energy of the
ejecta (Scenario B, see par. 4). Furthermore, both the density $n$
of the environment and the efficiency $\eta$ of the conversion of
kinetic energy into gamma rays should be moderately low. We assume
$n=5 \times 10^{-3}$ and $\eta\sim0.01$. In order to derive an
estimate of the energy produced by this burst, we look at the prompt
emission fluence and spectrum. GRB050319 prompt emission between 20
and 150~keV was fitted by a single powerlaw spectrum, with photon
index $\Gamma=2.1$ and had a fluence of $1.1\times10^{-6}$ ergs
cm$^{-1}$ (Cusumano et al. 2006). If we assume that the prompt
emission spectrum of this GRB is described by the Band function,
with spectral break below 20~keV and a typical low energy photon
index 1, we find that this burst emitted $7\times10^{52}$ ergs in
the 1-10000~keV band, on the basis of isotropic emission at redshift
z=3.24. Under the previous assumption on efficiency, density and
fraction of total energy which goes into the narrow outflow, a jet
break at 30~ks is compatible with a beaming angle of $\theta_N =
0.015$ rad.
% If the optical emission carries on without breaking up
%to $\sim400$~ks as in Panaitescu et al.~2006a, the lower limit on
%the opening angle of the wide component is $\theta_W > 0.012$ rad
%\textbf{Should I remove this statement? It may be useless to achieve
%our goal and confuse readers.} .
% Beaming angles of this
%order are not implausible, and  have been previously observed (e.g.
%in GRB980519, Jaunsen et al. 2001).

 We note that, strictly speaking, in Eq~\ref{9}, we should have
taken into account that the energy of the ejecta is increasing
during the afterglow. Nevertheless, considering the weak dependence
of $\theta$ on $E_{k,52}$, the value of $\theta$ we found can be
considered correct within a factor 2.

\begin{table*}
\begin{center}
\begin{tabular}{ccccccc}
\hline \\
GRB      & $\alpha_{O}$ &  $\alpha_{X,2}$ &$\beta_{X,2}$& $t_{X,2}$
(ks) & $\alpha_{X,3}$ & $\beta_{X,3}$ \\
\hline \\
050319  & $0.62\pm0.02$  & $0.48 \pm 0.03$     & $1.00\pm0.03$ & $29.93^+{2.55} _{-2.80}$ & $1.41^{+0.08} _{-0.07}$ & $1.12\pm0.07$  \\
050802  & $0.82\pm0.03$  & $0.63 \pm 0.03$     & $0.89\pm0.04$ & $5.0\pm0.3$              & $1.59\pm0.03$           & $0.88\pm0.04$\\
060605  & $0.83\pm0.04$  & $0.41 \pm 0.03$     & $1.04\pm0.07$ & $7.73\pm 0.38$           & $1.93^{+0.11} _{-0.10}$ & $1.20\pm0.09$  \\
 \hline
050401  &               & $0.56\pm0.02$           &$0.99\pm0.02$  & $4.27\pm0.52$                   & $1.44 \pm 0.07$ & $0.95\pm0.07$        \\
050607  &               & $0.54^{+0.09} _{-0.10}$ &$1.04\pm0.14$ & $16.2^{+6.4} _{-4.2}$   & $1.33^{+0.16} _{-0.11}$  &$1.17\pm0.20$ \\
050713A &               & $0.58\pm0.03$           &$1.27\pm0.04$ & $7.54^{+0.87}_{-0.80}$  & $1.21 \pm0.03         $  &$1.02\pm0.05$\\
 \hline
\end{tabular}

\caption{Results of the analysis of the bursts with chromatic breaks
considered in this paper. From left to right: burst name, decay
index in the optical, X-ray decay slope of the plateau phase, X-ray
spectral slope in the plateau phase, X-ray lightcurve break time,
X-ray late times decay index, X-ray late times spectral slope
}\label{tab1}
\end{center}
\end{table*}

\begin{table*}
\begin{center}
\begin{tabular}{cccccc}
\hline\hline
& & no injection & &  injection & \\

& $\beta$ & $\alpha $  &  $\alpha (\beta)$ & $\alpha$ & $\alpha (\beta)$  \\
\hline
ISM and spehrical expansion \\
\hline

$\nu_m<\nu<\nu_c$ & ${{p-1 \over 2}}$ (0.7) & ${3(p-1)\over4}$(1.05)
&
$\alpha={3\beta \over 2}$ & ${(2p-6)+(p+3)q \over 4}$ (0.38) &
$\alpha=(q-1)+\frac{(2+q)\beta}{2}$\\

$\nu>\nu_c$   &  ${{p\over 2}}$ (1.2)  &   ${3p-2 \over 4}$ (1.30) & $\alpha={3\beta-1 \over 2}$ & ${(2p-4)+(p+2)q\over 4}$ (0.75) & $\alpha=\frac{q-2}{2}+\frac{(2+q)\beta}{2}$ \\

& & & & &\\

\hline
ISM and jet expansion \\
\hline

$\nu_m<\nu<\nu_c$ & ${p-1 \over 2}$ (0.7)  & $p$ (2.4)
 &
$\alpha = 2\beta +1 $ & ${(2p-3) + (p+3)q \over 3}$ (1.5) &
$\alpha = {(4\beta-1) + 2(\beta+2)q \over 3}$ \\

$\nu>\nu_c$   & ${{p\over 2}}$ (1.2) & $p$ (2.4) & $\alpha=2\beta$ &
${2(p-1) + (p+2)q \over 3}$ (1.67)   & ${2(q-1) \over 3} +
{2(1+q)\beta \over3}$
\\
& & & & &\\
 \hline
Wind and spherical expansion \\
\hline
$\nu_m<\nu<\nu_c$   &  ${p-1\over 2}$ (0.7)  &   ${3p-1\over 4}$ (1.55)   &   $\alpha={3\beta+1 \over 2}$ & ${(2p-2)+(p+1)q \over 4}$ (1.13)& $\alpha=\frac{q}{2}+\frac{(2+q)\beta}{2}$\\

$\nu>\nu_c$   &  ${p\over 2}$ (1.2)  &   ${3p-2\over 4}$ (1.3)   &  $\alpha={3\beta-1 \over 2}$   &   ${(2p-4)+(p+2)q\over 4}$ (0.75) &  $\alpha=\frac{q-2}{2}+\frac{(2+q)\beta}{2}$\\

\end{tabular}
\caption{Table with the relations between the value of decay index
$\alpha$ and  the spectral slope $\beta$ in various afterglow models
with the inclusion of the cases of energy injection. The case of
$p<2$ is not included, and the self-absorption effect is not
discussed. We do not consider the case of observing frequencies
below $\nu_m$. The convention $F_\nu \propto t^{-\alpha}
\nu^{-\beta}$ is adopted here. The temporal indices with energy
injection are valid only for $q < 1$, and they reduce to the
standard case (without energy injection: Sari et al. 1998, Chevalier
\& Li 2000; with Energy injection: Zhang et al. 2006, Panaitescu et
al 2006b) when $q=1$. For $q>1$ the expressions with energy
injection are no longer valid, and the standard model applies. The
numerical values quoted in parentheses are for $p=2.4$ and $q=0.5$.
}\label{Table1}
\end{center}
\end{table*}
%\end{deluxetable}

\begin{table*}
\begin{center}
\begin{tabular}{cccccc}
\hline \\
                  & Fit at 20 ks    &                             & \hspace{1.5cm} & Fit at 70 ks & \\

\hline \\
Parameters        & Single powerlaw          & Broken powerlaw        & &    Single powerlaw               & Broken powerlaw\\
\hline \\

    $\beta_{1}$   & $0.88^{+0.04} _{-0.04}$  & $0.49\pm0.05$& &  $0.84 \pm0.05$                         & $0.58^{+0.19} _{-0.12}$ \\
    $E_{B}$       &                & $0.20^{+0.24} _{-0.14}$          & &                                         & $0.28^{+0.06} _{-0.05}$ \\
    $\beta_{2}$   &                & $0.99\pm0.05$          & &                                         & $1.08^{+0.19} _{-0.12}$ \\
    $E_{B-V}$     & $13.1^{+1.9} _{-1.6} \times 10^{-2}$&$4.1^{+3.0} _{-2.7}\times10^{-2}$ & & $8.80^{+0.45} _{-0.47} \times 10^{-2}$  & $4.0^{+2.9} _{-1.7} \times 10^{-2}$ \\
    $N_{H}$       & $<0.56$        & $0.62\pm0.20$                           & &  $<0.23$                                & $<0.87$ \\
    $\chi_{\nu}$  & $129.5/108$    & $110/107$                      & &                               $34.6/24$ & $22.2/23$ \\

\end{tabular}
\caption{Best fit values of the GRB050319 SED at 20 ks and 70 ks.
$N_{H}$ is expressed in units of $10^{22}$~cm$^{-2}$, the break
energy $E_{B}$ is given in keV, and the local extinction $E_{B-V}$
is in magnitudes. All upper limits are at 90\% c.l..} \label{tab2}
\end{center}
\end{table*}

\begin{table*}
\begin{center}
\begin{tabular}{cccccc}
\hline \\
                  & Fit at 20 ks    &                             & \hspace{1.5cm} & Fit at 70 ks & \\

\hline \\
Parameters        & Single powerlaw & Broken powerlaw & & Single powerlaw & Broken powerlaw\\
\hline \\

    $\beta_{1}$   & $0.86 \pm 0.02$& $0.89\pm0.04$               &           & $0.99\pm0.02$ \\
    $E_{B}$       &                & $4^{+5} _{-3} \times 10^{-3}$ &         &               \\
    $\beta_{2}$   &                & $1.39\pm0.04$               &           &               \\
    $E_{B-V}$     & $ 18 \pm 2 \times 10^{-2}$ & $18\pm0.02 \times 10^{-2}$  &  & $0.18$     \\
    $N_{H}$       & $0.26 \pm 0.04$& $0.29\pm0.04$               &           & $0.26$        \\
    $\chi_{\nu}$  & $120/104$      & $119/103$                   &           & $27/15$       \\

\end{tabular}
\caption{Best fit values of the GRB0500802 SED at 0.4-1 ks and 35-55
ks. $N_{H}$ is expressed in units of $10^{22}$~cm$^{-2}$, the break
energy $E_{B}$ is given in keV, and the local extinction $E_{B-V}$
is in magnitudes. In the case, the fit of the two SEDs was performed
by assuming a Galactic extinction law (all results are taken from
Oates et al. 2007).} \label{tab1a}
\end{center}
\end{table*}

\subsection{GRB050802}

In the case of  GRB050802, we only  briefly summarize the results
obtained by Oates et al.~(2007); the X-ray and optical lightcurves
are shown in Fig.\ref{fig4} (middle panel). The X-ray lightcurve
breaks from a decay slope of $\alpha_{X,2}=0.63\pm0.03$ to a slope
of $\alpha_{X,3} = 1.59\pm0.03$, $5.0\pm0.3$ ks after the trigger.
The optical lightcurve is well fitted by a single powerlaw decay
with slope $\alpha_{O} = 0.82\pm0.03$; the $3\sigma$ lower limit on
any possible break in the optical is t=$19$ks. Two SEDs were built
at 500s and 40ks after the trigger (Fig.~\ref{fig4}, middle panel).
In the case of GRB050802, the best fit was provided by adopting the
Gal model. Therefore, for this burst, the extinction was determined
by applying this law.
%{\bf For this particular burst, data are best
%reproduced by using a Galactic extinction law *** this should be
%clarified! Also, perhaps state why it was possible to disentangle?
%better data?**}.
 By applying the extinction determined in the early SED to the late time
SED, it was determined that the late UV/optical emission lies above
the extrapolated X-ray spectrum. This indicates that the optical
emission is not produced by the same outflow that is responsible for
the X-ray emission, regardless of where the synchrotron peak
frequency and cooling frequency lie. Instead, the double component
scenario described earlier was found to be consistent with the data
if the X-ray band lies below the synchrotron cooling frequency
$\nu_{C}$. In this case, with the values of parameters
$\alpha_{X,2}=0.63\pm0.03$ and $\beta_X = 0.88\pm0.04$ we can derive
$q = 0.51\pm0.06$
% the relation between $q$, $\alpha$ and $\beta$ is
%(Zhang et al. 2006)
%\begin{equation}
% q= (\alpha -\beta +1)/(1+\beta/2) \, .
%\label{q2}
%\end{equation}
%Since $\beta = 0.88\pm0.04$, this gives $q=0.51\pm0.03$.
 If the break at 5~ks is interpreted as  a jet break, the expected
post-break slope would be (see again Panaitescu et al.,~2006)
$\alpha = 2\beta +1 = 2.76\pm0.08$ in case the decay proceeds
without further energy injection, and $\alpha =1.83\pm0.17$
%\begin{equation}
% \alpha = 2\beta +1 - 0.67 (1-q)(\beta+2)
%\label{dal2}
%\end{equation}
in case there is no cessation of energy injection,
%It can be easily seen that Eq.~\ref{dal2} gives $\alpha= 1.92 \pm 0.13$,
which is consistent with the observed value of $\alpha_{X,3}$ within
$2\sigma$. Results of the analysis are shown in Tables \ref{tab1}
and \ref{tab1a}.

\subsection{GRB050922c}

\begin{figure*}
 \includegraphics[width=0.4\textwidth=0.8]{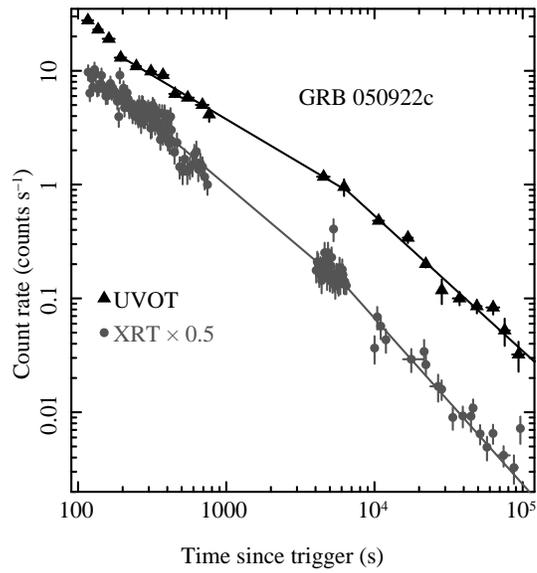}
 \caption{XRT and UVOT lightcurves of GRB050922c, the solid lines are the
best-fitting broken powerlaws.}\label{fig2}
\end{figure*}

A first inspection of GRB050922c data clearly shows a break in the
optical and XRT lightcurves (Fig. \ref{fig2}). In order to quantify
its significance, we fit the lightcurves with a single and a broken
powerlaw. The early optical emission shows some features
superimposed on the powerlaw decay, such as an evident bump at $\sim
150$s after the trigger. Therefore, we excluded from the fit UVOT
data taken during the first 200s after the trigger and, for
consistency, we did this with the X-ray data as well.
 In the case of the X-ray lightcurve, we found that the fit with an
unbroken powerlaw yields $\chi^{2}=219$ for 119 d.o.f., while a fit
with a broken powerlaw provides an $\chi^{2}=141$ for 117 d.o.f..
The Ftest (Bevington et al. 1969) indicates that the probability of
improvement by chance is less than $7\times10^{-12}$. For a broken
powerlaw, the best fit parameters are: initial decay slope
$\alpha_{X,2}=1.10\pm0.02$, break time $t_{X,b,2} =
6.45^{+1.83}_{-0.76}$ ks, and late decay slope $\alpha_{X,3} =
1.48^{+0.06}_{-0.04}$.
 For the optical lightcurve, a single powerlaw fit of the renormalised V,
B and U band lightcurves gives $\chi^{2} = 110.2$ for 18 d.o.f.,
whereas a broken powerlaw gives $\chi^{2} = 24$ for 16 d.o.f.. In
the latter case, the best fit parameters are
$\alpha_{O,1}=0.77\pm0.03$, $t_{O,b} = 6.23^{+1.16} _{-0.99}$ ks,
and $\alpha_{O,2}=1.20\pm0.05$. As we can see, from our reanalysis
and new  reduction of the X-ray and optical data, the break times in
the two bands turn out to be consistent with each other within
1$\sigma$, suggesting that the break in the X-ray lightcurve should
not be considered as achromatic, in contrast to what was suggested
by Panaitescu et al. (2006a).

\subsection{GRB060605}

The \textit{Swift} GRB060505 also shows a  canonical X-ray
lightcurve, with an initial steep decay, a shallow plateau and
finally a steep decay (Fig. \ref{fig4}, bottom panel). The decay
slopes of the three segments and the two break times are
$\alpha_{X,1} =  2.68^{+0.92} _{-0.52}$, $t_{X,1}=164.5^{+29.9}
_{-15.6}$s, $\alpha_{X,2}= 0.41\pm0.03$, $t_{X,2}=7.73\pm0.38$~ks,
$\alpha_{X,3}=1.93 ^{+0.11} _{-0.10}$. There is no evident strong
X-ray spectral evolution, since the X-ray energy index in the
plateau and in the steep decay are $\beta_{X,2}=1.04\pm0.07$ and
$\beta_{X,3} = 1.20\pm0.09$, consistent within $1\sigma$. In the
optical, GRB060605 shows a wide peak at few hundreds seconds after
the trigger, which is likely to be the beginning of the forward
shock emission (Oates et al. in prep.). In fitting the optical
lightcurve (Fig. \ref{fig4}, bottom panel), we considered all the
datapoints taken after 500s from the trigger. The single powerlaw
model provides a marginally acceptable fit, with $\chi^2=28$ for 13
d.o.f.. We then tried a broken powerlaw model, which gives a much
better fit with $\chi^{2}=10.6$ for 10 d.o.f.. The value of the late
decay slope is $\alpha_{O,2} = 3.3^{+\infty} _{-1.0}$, but it is not
well constrained; we can infer that it has a lower limit of $1.4$ at
95\% C.L.. The best fit values of the other parameters are
$\alpha_{O,1} = 0.85\pm0.04$ and $t_{O} = 23.5 ^{+5.9} _{-4.0}$ ks.
The 3$\sigma$ lower limit on the break time in the optical,
calculated as in the case of GRB050319, is $t_{O}=12.3$~ks. Ferrero
et al. (2008) present a dataset in which the optical afterglow is
well detected till $\sim1$ day after the trigger, and their data
show an evident break occurring 23.3~ks after the trigger, with a
late decay slope $\alpha_{O,2}=2.56\pm0.16$. We note that our best
fit values are consistent with those of Ferrero et al. (2008). Thus,
we can conclude that a break is present in the optical, but it is
inconsistent with $t_{X,2}$. Ferrero et al. (2008) suggest that the
different break times might be caused by some flaring activity in
the X-ray band that occurred around 6ks after the trigger. These
flares would have led to the conjecture of an X-ray afterglow
decaying shortly thereafter (see their paper for more details). We
will rather investigate the scenario in which GRB060605 has a
genuine chromatic break. For this GRB, we built up the SEDs at 5ks
and 20ks; the values of the best fit parameters are reported in
Table \ref{tab5}. As we can see, we cannot distinguish between the
single powerlaw and the broken powerlaw spectral fit on statistical
basis, since both models provide a similar reduced $\chi^2$.
However, an unbroken powerlaw model is ruled out by the fact that
X-ray and optical decay slopes are inconsistent at $7\sigma$ level
at 5~ks and $10\sigma$ level at 20~ks. We are thus left with a
scenario in which the spectrum is a broken powerlaw at 5~ks and
20~ks. Furthermore, we have to assume a wind circumburst environment
for the same reasons quoted for GRB050319. We reiterate that, in
this environment, the cooling frequency is supposed to increase.\\

If we fit the two SEDs with a broken powerlaw and restrict the break
energy between 0.005 and 1 keV, the low energy spectral indeces are
$\beta_{OX,E} = 0.54\pm0.07$, $\beta_{OX,L} = 0.71\pm0.09$, at 5 and
20~ks respectively. The break energy at 5ks is $0.008$~keV, with a
1~$\sigma$ positive error of $0.032$. This break energy value is
near the minimum allowed value of $0.005$~keV; we were not able to
find an 1$\sigma$ negative error.

% with a break at $\nu_{c} = 2.42 ^{+0.51} _{-0.40}$~keV.
%At 20ks, we found best fit values of $\beta_{OX,L} = 0.82 ^{+0.13}
%_{-0.08}$, $\beta_{OX,L}+0.5 = 1.32 ^{+0.13} _{-0.08}$, with the
%break at $1.45 ^{+0.22} _{-0.28}$~keV.
%The second set which can fit equally well the late SED is
%$\beta_{E,1}=0.70^{+0.08} _{-0.09}$, $\beta_{E,2}= \beta_{E,1} +0.5
%= 1.20^{+0.08} _{-0.09}$ and cooling break at $5.0^{+2.3} _{-1.2}$
%eV. We will refer to these sets as $S_{E,1}$, $S_{E,2}$, $S_{L,1}$
%and $S_{L,2}$ respectively. Since the cooling break must increase,
%we can rule out an evolution from $S_{E,1}$ to $S_{L,2}$, since the
%difference between the two cooling breaks is negative at $\sim
%4\sigma$ level. An evolution from $S_{E,2}$ to $S_{L,1}$ can be
%excluded because the energy indexes are inconsistent at more than
%$4\sigma$ level. The only evolutions left are between $S_{E,2}$ and
%$S_{L,2}$ and between $S_{E,2}$ and $S_{L,1}$ because they are
%consistent, although marginally, %with an increase of the cooling break with time.\\
 We note that the low energy spectral indices are consistent within
$2\sigma$. We assume an average low index $\overline{\beta}_{OX} =
0.60\pm0.06$ and a high energy index $\overline{\beta}_{OX} + 0.5 =
1.10\pm 0.06$ respectively. The first index has got to be that of
the Optical band. In the usual interpretation of the canonical X-ray
lightcurve, the break at 7.3ks corresponds to the end of energy
injection into the ejecta. If this is the right scenario, in a wind
density profile, the optical emission decay index should be higher
than that of the X-ray emission by 0.25. For example, we should
observe an optical decay slope $\alpha_O = \frac{3}{2}
\overline{\beta}_{OX} + 1/2 = 1.40 \pm 0.09$ after the end of the
injection; the X-ray flux decay index ought to be $\alpha_X =
\alpha_O - 0.25 = 1.15$. These predictions are clearly inconsistent
with the observed behaviour. The X-ray flux would decay faster than
1.15 if the cooling frequency moved above the X-ray band, but in
such a case the X-ray decay slope would be consistent with that of
the optical, which is inconsistent with observations, as stated
above.
% Even if the cooling frequency moved above the X-ray band,
%the X-ray band would decay faster, but the
% If, instead, the
%cooling break has transited above the X-ray band, this band and the
%optical band lie on the same spectral slope, and the decay should be
%the same in both bands.

% This spectral slope
%should be the weighted average between $\beta_{OX,E}$ and
%$\beta_{OX,L}$, that is $\overline{\beta}_{OX}=1.02\pm0.07$. In
%absence of energy injection the decay index of both components would
%be $\alpha  = 3/2 \overline{\beta}_{OX} + 1/2 = 2.04 \pm 0.11$. This
%value is consistent with $\alpha_{X,3}$, but totally inconsistent
%with the optical decay index till 20~ks after the trigger.
% In the second
%possible evolution, we note that the low energy spectral indices are
%consistent within $1\sigma$, as are the high energy slopes. We can
%therefore assume an average low and high energy index
%$\overline{\beta}_{1} = 0.97\pm0.08$ and $\overline{\beta}_{2} =
%1.47 \pm 0.08$ respectively. The first index corresponds to that of
%the optical band. In the canonical model, the break a 7.3ks in the
%X-ray lightcurve is associated to the end of an energy injection
%phase. If this were the right scenario, the optical emission should
%decay at a fast rate after the X-ray break, i.e. with a slope of
%$\alpha_{O} = 3/2 \overline{\beta}_{1} + 1/2 = 1.39 \pm 0.11$, which
%is inconsistent at $\sim 4\sigma$ level with the
%observed decay index.\\

 We try now to apply our model to interpret the behaviour of the X-ray
emission for this burst. Again, the idea is that the plateau,
extending till $7.3$ ks after the trigger, is due to forward shock
emission of ejecta expanding like they were spherical, with the
contribution of energy injection. For this burst, we suppose that
the X-ray band remains below the cooling frequency. In fact, by
using $\alpha_{X,2}$ and the weighted average energy index
$\overline{\beta}_{X} = 1.10 \pm 0.06$, from Tab.~\ref{Table1} we
derive $q=0.20 \pm 0.06$. Assuming that the end of the plateau phase
is due to a jet break with side expansion, the predicted decay slope
post break would be $\alpha = 3.20 \pm 0.12$ or $\alpha=1.48\pm0.20$
in case of cessation or continuation of the energy injection,
respectively (see Tab.\ref{Table1}). Again, the second value is
consistent with the observed result at $2\sigma$ level (
$\alpha_{X,3}=1.93 ^{+0.11} _{-0.10}$).
 In order to compute the opening angle $\theta_N$ of the outflow
responsible for the X-ray emission, we can follow the same procedure
as GRB050319 after estimating the total emitted energy. According to
Sato et al.(2006), the fluence in the 15-150 keV band of GRB060605
is $4.6\times10^{-7}$ erg cm$^{-2}$, while the spectrum is best
fitted by a simple powerlaw with photon index $\Gamma_{1} \sim
1.34$. Since this value suggests a high energy spectrum below the
break energy (Band et al. 1993), we can assume that the break energy
is occurring just above the BAT bandpass. Assuming that the high
energy photon index is $\Gamma_{2} = 2.3$ (the average value for
this parameter following Band et al. 1993) and redshift $z=3.8$, we
find that the isotropic energy emitted between 1 and 10000 keV is $E
\sim 3.5 \times 10^{52}$ ergs. The next step is to estimate the
kinetic energy of the ejecta and which fraction of it goes into the
narrow outflow. Now, in the case of GRB060605, a possible jet break
occurs in the optical not much later than the jet break in the
X-ray, Eq.~\ref{9} indicates that the opening angle $\theta_W$ of
the outflow responsible for the optical emission and $\theta_N$
could be close. Now, in our modelling (see section 4 for details),
it is intrinsically assumed that we have emissions from spherical
portions of two outflows, and the emitting surface of the narrow
outflow, responsible for the X-ray emission, is much less than the
surface of the wide outflow, which is producing the optical
emission. The approximation can hold if the beaming angles of the
two outflows are different enough. A way we can reconcile our
interpretation with the features of GRB060605 is by assuming that
the energy in the narrow component $E_N$ is much higher than the
energy $E_W$ carried by the wide outflow. In our theoretical
discussion, we have found that solutions with $E_N \simeq 30 E_W$
are possible (Scenario A'', see section 4). This solutions applies
in cases of density $n\simeq1$, and efficiency of conversion of
kinetic energy of the ejecta into $\gamma$-ray emission $\eta=0.2$.
We thus derive that the kinetic energy of the ejecta is $\sim
1.8\times 10^{53}$ ergs.  Now, if we apply this ratio of energies
and this density to GRB060605, then we derive, by using Eq.~\ref{9},
that the narrow outflow should have an opening angle $\theta_N =
0.019$ rad. The outflow responsible for the optical emission should
have $\theta_W = 0.05$.

\begin{table*}
\begin{center}
\begin{tabular}{ccccccc}
\hline \\
               &  Fit at 5 ks             & \hspace{1.5cm}          & & Fit at 20 ks & \\

\hline \\
Parameters     & Single Pow.              & Broken Pow.             & &   Single Pow.          & Broken Pow. \\
\hline \\
  $\beta_{1}$  & $1.01^{+0.07} _{-0.06}$  & $0.54\pm0.07$                   & & $1.16 \pm0.09$                     & $0.71\pm0.09$  \\
  $E_{B}$      &                          & $0.008^{+0.032}$                 & &                                  &$0.375 ^{+0.215} _{-0.350}$ \\
  $\beta_{2}$  &                          & $1.54\pm0.07$                   & &                                    & $1.21\pm0.09$  \\
  $E_{B-V}$    & $7.64^{+2.16} _{-0.64} \times 10^{-2}$  & $0.008^{+0.032}$          & & $12.5^{+0.5} _{-0.6} \times10^{-2}$& $<0.122$  \\
  $N_{H}$      & $<0.96$                  & $0.49^{+0.24} _{-0.24}$          & & $1.28^{+0.46} _{-0.43}$           & $1.48^{+0.71} _{-0.64}$  \\
  $\chi_{\nu}$ & 42.0/42                  & $41.4/41$                       & & $28.4/27$   & $26.4/26$ \\
  \hline
  \end{tabular}
\caption{Best fit values of the GRB060605 SED at 5 ks and 20 ks.
$N_{H}$ is expressed in units of $10^{22}$~cm$^{-2}$, the break
energy $E_{B}$ is given in keV, and the local reddening $E_{B-V}$ is
in magnitudes. Upper limits on column density and reddening are at
90\% c.l..} \label{tab5}
\end{center}
\end{table*}

\begin{table*}
\begin{center}
\begin{tabular}{cccc}
\hline \\
GRB & Observed $\alpha_{X,3}$ & Predicted $\alpha_{X,3}$ & $\theta_N$\\
\hline \\
050319 & $1.41^{+0.08} _{-0.07}$ & $1.31\pm0.10$ & 0.015 \\
050802 & $1.59\pm0.03          $ & $1.83\pm0.17$ & 0.017 \\
060605 & $1.93^{+0.11} _{-0.10}$ & $1.48\pm0.20$ & 0.019 \\
\hline \\
050401  & $1.44 \pm 0.07$         & $1.74\pm0.09$  & 0.006 \\
050607  & $1.33^{+0.16} _{-0.11}$ & $1.57\pm0.41$  & 0.023 \\
050713A & $1.21 \pm 0.03$         & $1.44\pm0.11$  & 0.008 \\
 \hline
 \end{tabular}

\caption{GRBs with chromatic breaks considered in this work. The
table shows the late decay slope observed in the X-ray, the slope
predicted by our model, and the inferred values of the beaming angle
for the narrow outflow. In the case of GRB050802, values are taken
by Oates et al.~(2007).} \label{tab8}

\end{center}
\end{table*}

\subsection{GRBs with X-ray data analysis only}

 All the bursts for which we built the optical and X-ray SEDs have
their redshift known by spectroscopy, while the following other
objects in our sample do not have known redshifts (except 050401).
However, since they are studied in the X-ray band only, the lack of
a redshift basically does not affect our results and conclusions.

GRB050401 - A break is evident in the X-ray lightcurve of this GRB:
the decay slope changes from $\alpha_{X,2} = 0.56 \pm 0.02$ to
$\alpha_{X,3} = 1.44 \pm 0.07$ at $t_{X,2}= 4.27\pm0.52$ ks. There
is not strong spectral evolution throughout the whole observation,
since the spectral index is always consistent with
$\beta_{X}=0.99\pm0.02$. Again, if the X-ray band is below
$\nu_{C}$, then the energy injection parameter would be $q = 0.39
\pm 0.03$. If the outflow responsible for the X-ray emission
underwent a jet break without energy injection, the predicted slope
of the flux decay would be $\alpha = 2.98 \pm 0.04$, which is
inconsistent with the value we observe. However, in the presence of
energy injection the predicted value is $\alpha=1.74\pm0.09$, which
is consistent with the observed X-ray decay slope at $\sim2.3\sigma$
level. In order to compute the beaming angle of the outflow
responsible for the X-ray emission, we need to make some
assumptions. We will assume that the Energy of narrow outflow
responsible for the X-ray emission is $10\%$ that of the a wider
outflow that produces the optical emission (Scenario B), and an
efficiency $\eta=0.01$ and a density $n=5\times 10^{-3}$. We have
$E_{\gamma}=3.5 \times 10^{53}$~ergs (Golenetskii et al. 2005). With
these assumptions for density, efficiency and ratios of kinetic
energies the jet beaming angle of the narrow component turns out to
be $\theta_N=0.006$ rad (Eq.~\ref{9}).

 GRB050607 - This burst exhibits an evident break in the X-ray
lightcurve, since its decay slopes change from $\alpha_{X,2} = 0.54
^{+0.09} _{-0.10}$ to $\alpha_{X,3} = 1.33 ^{+0.16} _{-0.11}$ at
$16.2^{+6.4} _{-4.2}$~ks. The X-ray spectrum does not show evidence
of evolution at the break time and has an average energy index of
$\beta_{X} = 1.07 \pm 0.11$. Assuming that the X-ray band is above
the cooling frequency, the values of $\beta_X$ and $\alpha_{X,2}$
imply $q = 0.59 \pm 0.23$ (Tab.~\ref{Table1}). Without late time
energy injection, the subsequent jet decay slope would be $\alpha =
2.14 \pm 0.22$, while with energy injection the predicted value is
$\alpha = 1.57 \pm 0.41$. The latter is consistent with the observed
value of $\alpha_{X,3}$, within $\sim1\sigma$. In order to derive
the beaming angle of the narrow outflow, we need an estimate of the
burst energetics. Since the redshift of this burst is presently
unknown, we adopted $z=2.5$ (i.e. about the average Swift GRB
redshift, Jakobsson et al. 2006) and a prompt emission spectral
index  estimated by the Band function, with a high energy photon
index of $\sim$ 2.1 in the energy band from 15 keV  to 10000 keV and
of $\sim$1 below 15~keV (Pagani et al. 2006 report $1.83\pm0.14$ in
the range 15-150 keV). Under this hypothesis, the energy emitted by
the burst would be $\sim 2.4\times10^{52}$~ergs. We can assume that
90\% of the kinetic energy of the outflows is carried by the broad
one, and we can take $\nu_c$ below the X-ray band (scenario B,
section 4); other assumptions are $n=5\times 10^{-3}$, $\eta=0.01$.
With these hypothesis in place, we obtain a beaming angle of
$\theta_N=0.023$ rad.

 GRB050713A - the X-ray lightcurve of this burst shows a break at
$t_{X,2}=7.54^{+0.87}_{-0.80}$ ks, after which the decay slope
increases from $\alpha_{X,2} = 0.58 \pm 0.03$ to $\alpha_{X,3} =
1.21 \pm 0.03$. The spectral index, throughout the whole
observation, is $\beta_{X}=1.17 \pm0.03$. The energy injection
parameter, again for the case of X-ray band above $\nu_{C}$, is
$q=0.38 \pm 0.06$. The expected slope at late times would be
$\alpha=2.34\pm0.06$ or $\alpha=1.44\pm0.11$ in case of cessation or
continuation of the energy injection process, respectively. The
latter is consistent with the observed decay slope in the X-ray band
within 2$\sigma$. To calculate the beaming angle of the narrow
outflow, we made again an assumption on the (currently unknown)
burst redshift. By using $z=2.5$, and taking the values of fluence
and spectral parameters published in Morris et al. (2007), we infer
an isotropic $\gamma$-ray energy of $E_{\gamma} \sim 2.5 \times
10^{53}$~ergs. With the same assumptions made for GRB050607, we
obtain $\theta_N=0.008$ rad.

 \section{Discussion}
\label{disc} Results reported in the previous section show that a
single outflow model cannot explain the behaviour of the GRBs with
chromatic breaks we have considered. Instead, we found that if the
X-ray flux is attributed to ejecta which are decoupled from those
responsible for the optical, the observed behaviours of these GRBs
can be explained. In the theoretical modelling of GRBs, a double
component outflow has already been put forward, even before the
launch of Swift (e.g. Berger et al. 2003, Peng et al. 2005). It has
been invoked to explain the complex temporal behaviour of X-ray and
optical emissions of the exceptional GRB080319B (Racusin et al.
2008). In this section, we would like to explore the viability of
the two-component jet model with the important addition of a
continuous energy injection, from a theoretical point of view.

%In this section, we explore the viability of our model from a
%theoretical point of view.
 The basic picture is based on ejecta with two different degrees
of collimation. The narrow outflow generates the X-ray emission,
while the wide one the optical.  Both emissions are due to the usual
forward shock, which has a synchrotron spectrum consisting of
powerlaws connected at particular frequencies (Sari, Piran \&
Narayan 1998), i.e. the synchrotron frequency $\nu_{m}$ and the
cooling frequency $\nu_{c}$. In this paper, we use the expressions
of $\nu_{m}$, $\nu_{c}$ and of the peak flux $F_{max}$ as determined
in Zhang et al. (2007) for a constant density medium:

\begin{eqnarray}\label{array1}
  \nonumber F_{max} &=& 1600 (1+z) D_{28}  \epsilon_{B,-2} ^{1/2}
E_{K,52} n^{1/2} \, {\rm \mu Jy} \\
  \nonumber\nu_{m} &=&  3.3 \times 10^{12} \left (\frac{p-2}{p-1} \right )
^{2}
\left (1+z \right )^{1/2} \epsilon_{B,-2} ^{1/2} \epsilon_{e,-1} ^{2} \\
\nonumber & & \times
E_{K,52} ^{1/2} t_{d} ^{-3/2} \, {\rm Hz} \\
  \nonumber\nu_{c} &=&  6.3 \times 10^{15} (1+z)^{-1/2}
\epsilon_{B,-2}^{-3/2} E_{K,52} ^{-1/2} n^{-1} t_{d}^{-1/2}  \, {\rm Hz}\, , \\
\end{eqnarray}
where  $z$ is the redshift, $D_{28}$ is the luminosity distance in
units of $10^{28}$~cm, $\epsilon_{B,-2}$ and $\epsilon_{e,-1}$ are
the ratios between the magnetic/electron and kinetic energy of the
ejecta (in units of $10^{-2}$ and $10^{-1}$ respectively), $E_{52}$
is the isotropic kinetic energy as measured in the observer rest
frame and normalized to $10^{52}$~ergs, $n$ is the particle density
in cm$^{-3}$, $p$ is the index of the the powerlaw energy
distribution of radiating electrons, and $t_d$ is the observer time
in days.

\begin{figure*}
 \includegraphics[width=1\textwidth=0.8]{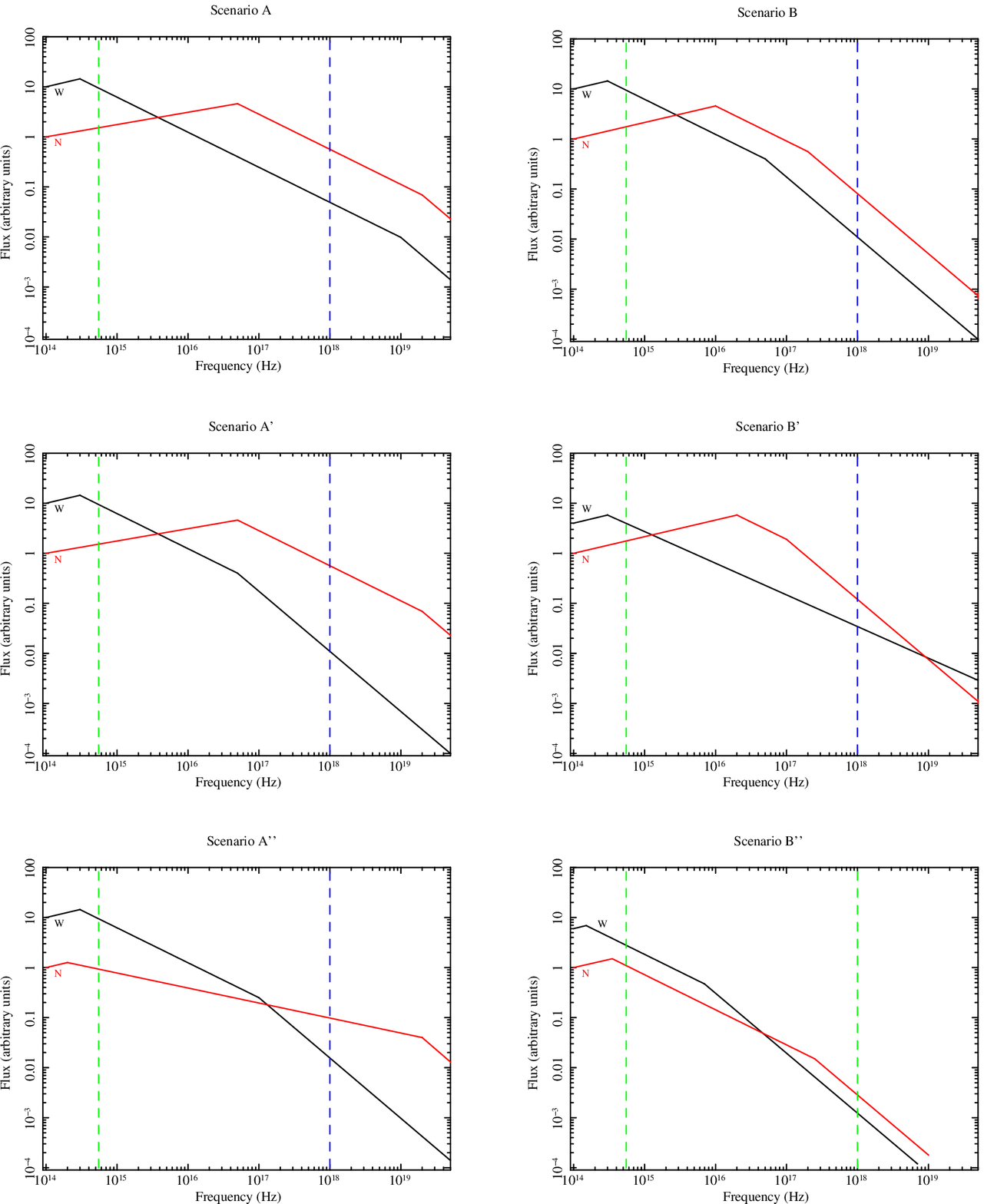}
 \caption{The 6 configurations of the wide (W) and narrow (N) components explored in the text.
 Vertical lines indicate the optical and X-ray band respectively.}\label{scenarios}
\end{figure*}

By taking $z=2.5$ (as for an average \textit{Swift} GRB, see
previous sections) and a cosmology with $H_{0}=71, \Omega=0.3,
\Lambda=0.7$, gives $D_{28} = 6.2$. We adopt a typical value of
$p=2.4$, which gives an energy index between $\nu_{m}$ and $\nu_{c}$
of $\beta = (p-1)/2 = 0.7$. Below  $\nu_{m}$ we assume a standard
synchrotron spectrum rising with  $\beta=-1/3$. In order to take
into account the energy injection, we assume that the luminosity of
the GRB central engine scales as $L \propto t^{-q}$, with a typical
value of $q = 0.5$ (Zhang et al. 2006). This corresponds to an
increase of kinetic energy of the ejecta of the kind $E \propto
t^{(1-q)} = t^{0.5}$. All these assumptions allow us to recalculate
the coefficient in the formulae of ~\ref{array1} and change the time
dependencies, taking into account the increase in energy. We obtain

 \begin{eqnarray}\label{eqnset1}
    \nonumber F_{max} &=& 2.55\times10^{3} \epsilon_{B,-2} ^{1/2}
E_{52,0} n^{1/2} t_{d} ^{1/2} \, {\rm \mu Jy} \\
    \nonumber \nu_{m} &=& 2.1 \times10^{12}
\epsilon_{B,-2} ^{1/2} \epsilon_{e,-1} ^{2} E_{52,0} ^{1/2} t_{d}
    ^{-5/4} \, {\rm Hz} \\
    \nu_{c} &=& 4.4 \times10^{14} \epsilon_{B,-2} ^{-3/2}
    E_{52,0} ^{-1/2} n^{-1} t_{d} ^{-3/4} \, {\rm Hz} \,
 \end{eqnarray}
where $E_{52,0}$ is the isotropic kinetic energy at 300s after the
trigger. We chose this time because it is typically from 300s that
the slow decline phase is observed in both the X-ray and optical
afterglows. Furthermore, we require our scenario to work up to 0.1
days after trigger, since it is typically around $\sim0.1$ days that
the plateau phase ends. To distinguish between the narrow and wide
component, we use the pedices ``n'' and ``w'' respectively, while
``O'' and ``X'' indicate the optical and X-ray band. For the optical
and X-ray frequencies, we used the values $\nu_O =
5.5\times10^{14}$~Hz and $\nu_X = 10^{18}$~Hz, respectively.
Therefore, for instance, $f_{O,w}$ is the optical flux due to the
wide component.\\ In the following treatment, we shall be discussing
six possible scenarios. In order for our model to work, the
narrow/wide component should not contribute significantly to the
optical/X-ray flux. We translate this ``condition of
non-interference'' by requiring that the optical flux of the narrow
component is at maximum one half of that of the wide one, and a
similar condition for the X-ray band. The six different scenarios we
are considering reflect six different possible hierarchies between
the various frequencies. Scenarios A and B deal with the case in
which both $\nu_{c,n}$ and $\nu_{c,w}$ lie above or below the X-ray
band, respectively. The next two cases, A' and B', are a variant of
the previous ones, in which $\nu_{c,w}$ and $\nu_{c,n}$ do not lie
on the same side with respect to the X-ray frequency. Cases A'' and
B'' show the same arrangements of frequencies as A' and B', but the
synchrotron peak frequency of the narrow component is below the
optical since the beginning of observations. Our data do not allow
to distinguish between the cases A, A', and A'' (or B, B' and B'').
We require $\nu_{m,w} < \nu_O$ and $\nu_{m,n} < \nu_X$, consistently
with the absence of an increase in the optical and X-ray flux at
early times in the datasets we have analyzed. All scenarios are
summarized in Fig.~\ref{scenarios}.
% We assume
%$\nu_{m,n}> \nu_O$ or $\nu_{m,n} < \nu_O$ since in this case the non
%interference condition in the optical band is more easily satisfied.
% We require $\nu_{m,w} < \nu_O$ and $\nu_{m,n} <
%\nu_X$, consistently with the absence of an increase in the optical
%and X-ray flux at early times in the datasets we have analyzed.
% All these relative conditions between the various
%frequencies are required to hold for the whole duration of the
%observation.
% In the following, while developing our scenarios we
%will assume they hold from 300s to 0.1d after the trigger, since in
%this interval all our GBRs lightcurves are well sampled.
In \S~\ref{sum}, we discuss the extension of validity of the
conditions we pose after 0.1~d.

\subsection{Scenario A}

 The conditions to apply in scenario A are:
\begin{equation}\label{100}
    f_{O,w} > 2 f_{n,O} \qquad {\rm at \, 0.1 \, d \, after \, the
\, trigger.}
\end{equation}

\begin{equation}\label{200}
   f_{X,w} < \frac{1}{2} f_{n,w}
\end{equation}

\begin{equation}\label{1}
  \nu_{m,w} < \nu_O \qquad {\rm at \, 300 \, s \, after \, the
\, trigger.}
\end{equation}

\begin{equation}\label{1b}
  \nu_{m,n} > \nu_O \qquad {\rm at \, 0.1 \, d \, after \, the
\, trigger.}
\end{equation}

\begin{equation}\label{2}
 \nu_{X} < \nu_{c,w} \qquad {\rm at \, 0.1 \, d \, after \, the
\, trigger.}
\end{equation}

\begin{equation}\label{2b}
 \nu_{X} < \nu_{c,n} \qquad {\rm at \, 0.1 \, d \, after \, the
\, trigger.}
\end{equation}

\begin{equation}\label{3}
 \nu_{m,n} < \nu_{X} \qquad {\rm at \, 300 \, s \, after \, the
\, trigger.}
\end{equation}
It is easy to verify that, if the above conditions are satisfied at
the time indicated, they are also valid for the whole interval in
which we are interested, i.e. between 300s and 0.1 days after the
trigger. Since the second condition describes the evolution of the
flux below the cooling frequencies for both components, time
dependencies cancel out.  The first condition (Eq.~\ref{100}) can be
written as

\begin{eqnarray}
 \epsilon_{B,-2,w}^{1/2} E_{52,0,w}
  \left (\frac {5.5\times10^{14}} {2.1\times 10^{12}
 \epsilon_{B,-2,w}^{1/2} \epsilon_{e,-1,w} ^{2} E_{52,0,w}
 ^{1/2} 0.1^{-5/4} } \right )^{-0.7} \\
> \nonumber 2 \epsilon_{B,-2,n} ^{1/2} E_{52,0,n}
 \left (\frac { 5.5\times10^{14}}{2.1 \times 10^{12}
 \epsilon_{B,-2,n}^{1/2} \epsilon_{e,-1,n} ^{2} E_{52,0,n}
 ^{1/2} 0.1^{-5/4}} \right )^{1/3}
\end{eqnarray}
which, after some iterations, can be rearranged into

\begin{equation}\label{100b}
 \epsilon_{B,-2,w} ^{0.85} E_{52,0,w} ^{1.35} \epsilon_{e,-1,w}
 ^{1.4} > 29.5 \epsilon_{B,-2,n} ^{1/3} E_{52,0,n} ^{5/6}
 \epsilon_{e,-1,n}^{-2/3}\, .
\end{equation}

Similarly, the second condition (Eq.~\ref{200}) can be expressed as

\begin{eqnarray}
\epsilon_{B,-2,w}^{1/2} E_{52,0,w} \left (\frac{10^{18}}
{2.1\times10^{12} \epsilon_{B,-2,w} ^{1/2} \epsilon_{e,-1,w} ^{2}
E_{52,0,w} ^{1/2} } \right ) ^{-0.7} \\
\nonumber  < \frac{1}{2} \epsilon_{B,-2,n}^{1/2} E_{52,0,n} \left
(\frac{10^{18}}{2.1\times10^{12} \epsilon_{B,-2,n} ^{1/2}
\epsilon_{e,-1,n} ^{2} E_{52,0,n} ^{1/2} } \right ) ^{-0.7}
\end{eqnarray}
which simplifies to

\begin{equation}\label{200b}
\epsilon_{B,-2,w} ^{0.85} E_{52,0,w} ^{1.35} \epsilon_{e,-1,w}
^{1.4} < \frac{1}{2} \epsilon_{B,-2,n} ^{0.85} E_{0,52,n} ^{1.35}
\epsilon_{e,-1,n}^{1.4}  \, .
\end{equation}

From these two inequalities we have

\begin{equation}\label{4}
E_{52,0,n} > 2.73\times10^{3}\epsilon_{B,-2,n}^{-1}
\epsilon_{e,-1,n} ^{-4}\, .
\end{equation}

We can now obtain a constraint on $E_{52,0,n}$ from Eq.~\ref{2b}

\begin{equation}\label{3d}
 E_{52,0,n}  < 6.1 \times 10^{-6} \epsilon_{B,-2,n} ^{-3} n^{-2} \, ,
\end{equation}
which,  substituted in Eq.~\ref{4}, gives

 \begin{equation}\label{3e}
\epsilon_{B,-2,n} < 4.8 \times 10^{-5} \epsilon_{e,-1,n} ^2 n^{-1}
\, .
 \end{equation}
% Now, we write Eq.~\ref{2b} in terms of physical parameters involved:
%
% \begin{equation}\label{2c}
%\epsilon_{B,-2,n}^{1/2} E_{52,0,n}^{1/2} \epsilon_{e,-1,n}^{2} >
%14.7
% \end{equation}

By substituting Eq.~\ref{3e} into Eq.~\ref{4} and after some
manipulating, we have

 \begin{equation}\label{1c}
E_{52,0,n} > 5.4 \times 10^{7} \epsilon_{e,-1,n} ^{-6} n \, .
 \end{equation}

%\textbf{Silvia, Mat: I think I've found a more constraining
%relation. By substituting Eq. \ref{4b} into \ref{4}, we obtain}

%\begin{equation}
%E_{52,0,n} > 6.3 \times 10^{6} \epsilon_{e,-1,n} ^{-6} n
%\end{equation}

%\textbf{however, this doesn't change the results of this
%subsection.}

From Eq.~\ref{1c}, we can infer that the value of
$\epsilon_{e,-1,n}$ must be very high, in order to avoid an
unreasonable value for the energy of the narrow outflow. By assuming
$\epsilon_{e,-1,n}=3.3$, i.e. the maximum value (which is given at
equipartition), we obtain $E_{52,0,n} > 4 \times 10^{4} n$. With
this value of $\epsilon_{e,-1,n}$, a constraint on $\epsilon_{B}$
can now be obtained from Eq~\ref{3e}; by assuming $n=0.01$ it gives
$\epsilon_{B,-2,n} < 5 \times 10^{-2}$, which is a very low value.

We try now to derive some constraints on the physical parameters of
the wide component. By solving Eq.~\ref{100b} for the parameter
$E_{52,0,n}$ and substituting it into Eq.~\ref{200b}, we derive

\begin{equation}\label{300}
E_{52,0,w}^{-0.79} \epsilon_{B,-2,w} ^{-0.5}
\epsilon_{e,-1,w}^{-0.8} < 2.3\times 10^{-3} \epsilon_{B,-2,n}
\epsilon_{e,-1,n} ^{2.35} \, .
 \end{equation}
which can be combined with Eq.~\ref{2}
\begin{equation}\label{2c}
E_{52,0,w}^{-1/2} \epsilon_{B,-2,w}^{-3/2} n^{-1} > 404.4
\end{equation}
to obtain
\begin{equation}\label{400}
E_{52,0,w} > 4 \times 10^6 n^{0.54} \epsilon_{e,-1,w} ^{-1.29}
\epsilon_{B,-2,n} ^{-0.47} \epsilon_{e,-1,n} ^{-3.8} \, .
\end{equation}

Under the previous assumption of $\epsilon_{e,-1,n}=3.3$ and taking
$\epsilon_{B,-2,n}= 10^{-2}$, we derive  $E_{52,0,w} > 3.5 \times
10^{4} \epsilon_{e,-2,w}^{-1.29}$. Again, the fraction of the energy
given to the electrons must be close to equipartition, in order to
avoid very high values of the energy of the wide component. If
$\epsilon_{e,-2,w} \sim 3.3$, we obtain $E_{52,0,w} > 8 \times
10^{3}$. It is physically implausible to have a value of
$E_{52,0,w}$ much higher than this lower limit. Apart from these
caveats, we  can now show that the set of inequalities assumed
within scenario A cannot be simultaneously verified. In fact,
Eq.~\ref{1} reads

\begin{equation}\label{500}
\epsilon_{B,-2,w}^{1/2} \epsilon_{e,-1,w}^2 E_{52,0,w} ^{1/2} <
0.221 \, .
\end{equation}

With  the values we just obtained for $\epsilon_{e,-1,w}^2$ and
$E_{52,0,w}$, Eq.~\ref{500} requires an extremely small
value of the magnetic energy, $\epsilon_{B,-2,w} <
10^{-9}$.
% We notice that Eq ~\ref{100b} and ~\ref{200b} have the same left
%hand member which, in ~\ref{100b} must be greater than a certain
%quantity, which depends on $E_{52,0,n}$, $\epsilon_{e,-1,n}$ and
%$\epsilon_{B,-2,n}$. In ~\ref{200b}, the same left hand member must
%be smaller than another quantity, again dependent on $E_{52,0,n}$,
%$\epsilon_{e,-1,n}$ and $\epsilon_{B,-2,n}$.
On the other hand, by substituting the lower limits on $E_{52,0,n}$,
$E_{52,0,w}$ and $\epsilon_{e,-1,n} = \epsilon_{e,-1,w}^2 = 3.3$
into Eq.~\ref{100b} and ~\ref{200b}, we can derive the following
inequalities

\begin{equation}\label{600} 8.8 \times10^{3} \epsilon_{B,-2,n}^{0.85} >
10 \times 10^{6} \epsilon_{B,-2,w} ^{0.85} > 2 \times 10^{3}
\epsilon_{B,-2,n} ^{1/3} \, , \end{equation}

and, in turn, a  lower limit on
$\epsilon_{B,-2,n}>7.7\times10^{-3}$. By substituting this value in
the right member  of Eq.~\ref{600} gives $\epsilon_{B,-2,w} > 3.3
\times 10^{-3}$, which is in contradiction with what was derived
from Eq.~\ref{500}.

\subsection{Scenario A'}

We now consider a variant of the previous scenario, in which the
cooling frequency of the wide component lays below the X-ray
frequency but above the optical band. As already mentioned, our
model cannot distinguish between this scenario and that described in
the previous subsection. The set of  conditions expressed in
Eqs.~\ref{100} - \ref{3} are modified as follows:

\begin{equation}\label{a2}
    f_{O,w} > 2 f_{O,n} \qquad {\rm at \, 0.1 \, d \, after \, the
\, trigger.}
\end{equation}

\begin{equation}\label{a1}
   f_{X,w} < \frac{1}{2} f_{X,n} \qquad {\rm at \, 300 \, s \, after \, the
\, trigger.}
\end{equation}

\begin{equation}\label{a7}
  \nu_{m,w} < \nu_O  \qquad {\rm at \, 300 \, s \, after \, the
\, trigger.}
\end{equation}

\begin{equation}\label{a8}
\nu_{O} < \nu_{c,w} \qquad {\rm at \, 0.1 \, d \, after \, the \,
trigger.}
\end{equation}

\begin{equation}\label{a4}
  \nu_{x} > \nu_{c,w}  \qquad {\rm at \, 300 \, s \, after \, the
\, trigger}
\end{equation}

\begin{equation}\label{a5}
  \nu_{x} > \nu_{m,n}  \qquad {\rm at \, 300 \, s \, after \, the
\, trigger}
\end{equation}

\begin{equation}\label{a6}
  \nu_{m,n} > \nu_{O}  \qquad {\rm at \, 0.1 \, d \, after \, the
\, trigger}
\end{equation}

\begin{equation}\label{a3}
  \nu_{x} < \nu_{c,n}  \qquad {\rm at \, 0.1 \, d \, after \, the
\, trigger \, .}
\end{equation}

Notice that Eq.~\ref{a6} must now hold at 0.1 days after the
trigger, while Eq.~\ref{a1} must be satisfied at 300s. These conditions
are translated into the following inequalities:

\begin{equation}\label{b}
\epsilon_{B,-2,w} ^{0.85} E_{52,0,w} ^{1.35} \epsilon_{e,-1,w}^{1.4}
> 29.5 \epsilon_{B,-2,n} ^{1/3} E_{52,0,n} ^{5/6}
\epsilon_{e,-1,n}^{-2/3}\, ,
\end{equation}

\begin{equation}\label{a}
\epsilon_{B,-2,w} ^{0.1} \epsilon_{e,-1,w} ^{1.4} E_{52,0,w} ^{1.1}
< 2.85 \epsilon_{B,-2,n} ^{0.85} \epsilon_{e,-1,n} ^{1.4}
E_{52,0,n}^{1.35} n^{1/2}
\end{equation}

\begin{equation}\label{a100}
\epsilon_{B,-2,w} ^{1/2} \epsilon_{e,-1,w} ^{2} E_{52,0,w} ^{1/2} <
0.22
\end{equation}

\begin{equation}\label{a200}
 \epsilon_{B,-2,w}^{-3/2} E_{52,0,w} ^{-1/2} n^{-1} > 0.22
\end{equation}

\begin{equation}\label{a300}
\epsilon_{B,-2,w} ^{-3/2} E_{52,0,w} ^{-1/2} n^{-1} < 32.5
\end{equation}

\begin{equation}\label{a400}
\epsilon_{B,-2,n} ^{1/2} \epsilon_{e,-1,n} ^{2} E_{52,0,n} ^{1/2} <
4\times 10^{2}
\end{equation}

\begin{equation}\label{a500}
\epsilon_{B,-2,n} ^{1/2} \epsilon_{e,-1,n} ^{2} E_{52,0,n} ^{1/2}
>14.7
\end{equation}

\begin{equation}\label{a600}
\epsilon_{B,-2,n}^{-3/2} E_{52,0,n} ^{-1/2} n^{-1} > 4 \times 10^{2} \, .
\end{equation}

By combining Eq.~\ref{a100} with Eq.~\ref{a300} we obtain

\begin{equation}\label{a800}
E_{52,0,w} < 0.35 \epsilon_{e,-1,w} ^{-6} n \, .
\end{equation}
 In order not to restrict ourselves to solutions with very low
$E_{52,0,w}$, we will assume a quite small value of
$\epsilon_{e,-1,w}$

% According to Frail et al. (2001; see further for more details),
%.$E_{52,0,w}$ should be as high as $10-100$, which in turn requires
%a quite small  value of $\epsilon_{e,-1,w}$.

Rearranging Eq.~\ref{a300}, we derive a constraint on
$\epsilon_{B,-2,w}$:

\begin{equation}\label{a900}
\epsilon_{B,-2,w} >0.10 E_{52,0,w}^{-1/3} n^{-2/3} \, ,
\end{equation}
% This last equation indicates that $\epsilon_{B,-2,w}$
%must be of the order $\sim0.5$ or higher.
%Similarly, from Eq.~\ref{a300} we can infer that
%By explicating $\epsilon_{e,-1,w}$ from Eq.~\ref{b} and placing it
%into Eq.~\ref{a}, we obtain the following inequality:
%\begin{equation}\label{a800}
%\epsilon_{B,-2,w} ^{-0.75} E~_{52,0,w} ^{-0.25} < 18.5
%\epsilon_{e,-1,n} ^{2.07} \epsilon_{B,-2,n} ^{0.52},
%E_{52,0,n}^{0.52} n^{0.5}
%\end{equation}
while  solving Eq.~\ref{a500} for $\epsilon_{B,-2,n}$ and
substituting it into Eq.~\ref{a600}, we obtain

\begin{equation}\label{a950}
E_{52,0,n} > 1.28 \times 10^{6} \epsilon_{e,-1,n}^{-6} n  \, .
\end{equation}

 From this last equation we infer that $\epsilon_{e,-1,n}$ cannot be
very far from the maximum value of 3.3, achieved at equipartition,
to avoid extremely high values of energy of the narrow outflow. If
$\epsilon_{e,-1,n}=3.3$, then $E_{52,0,n}
>10^3 n$.

Finally, from Eq.~\ref{a600} we
obtain a constraint on $\epsilon_{B,-2,n}$

\begin{equation}\label{a1000}
\epsilon_{B,-2,n} < 0.019 E_{52,0,n} ^{-1/3} n^{-2/3} \, .
\end{equation}

%Another constraint can be obtained by substituting the Energy from
%Eq.~\ref{600} and Eq.~\ref{500} and forcing $\epsilon_{e,-1,n}=3.3$:

%\begin{equation}\label{a1000b}
%\epsilon_{B,-2,n} < 0.0018 n^{-1}
%\end{equation}

%If the values of density are not extremely small, this condition is
%more constraining than the \ref{a1000}.

%this equation implies that $\epsilon_{B,-2,n}$ must be very small,
%less than a few thousandth; this limit does not depend much on the
%value $E_{52,0,n}$.

Now we will show that the previous set of inequalities can be solved
simultaneously by assuming not unreasonable values of the physical
parameters, provided that we limit ourselves to a scenario in which
the circumburst density is relatively small, $n \sim 10^{-3}$, which
makes the set of conditions easier to meet. Let us first assume that
$\epsilon_{e,-1,n}=3.3$ and that the energy of the narrow component
is, $E_{52,0,n} = 5$.
%As will be clearer in the
%following, a low value of $E_{52,0,n}$ permits a low and more
%physically plausible value for $E_{52,0,w}$.
For the chosen values
of these two parameters, Eq.~\ref{a500} requires $\epsilon_{B,-2,n}>0.6$.
Taking $\epsilon_{B,-2,n}=1$, from Eq.~\ref{a100} and Eq.~\ref{b}
we can derive a lower limit on the values of $E_{52,0,w}$.
In fact, Eq.~\ref{b} becomes

\begin{equation}\label{a1100}
\epsilon_{B,-2,w}^{0.85} \epsilon_{e,-1,w}^{1.4} E_{52,0,w}^{1.35}
> 50 \, ,
\end{equation}
which, combined with Eq.~\ref{a100}, gives

\begin{equation}\label{a1200}
\epsilon_{B,-2,w}^{0.5} E_{52,0,w} > 150 \, .
\end{equation}
Now, the highest value possible of $\epsilon_{B,-2,w}$ is reached at
the equipartition, $\epsilon_{B,-2,w} = 33$; in such a case,
$E_{52,0,w} > 25$. By assuming a more reasonable value of
$\epsilon_{B,-2,w} = 10$ (Panaitescu \& Kumar 2001b, Yost et al.
2003) give instead $E_{52,0,w} > 50$.

%Eq.\ref{a1200}
%tells us the bigger the energy of the narrow component, the higher
%the energy of the wide component, because the right member of Eq.
%~\ref{a1200} is proportional to $E_{52,0,n}$. Eq.~\ref{a1200} cannot
%change very much by choosing different values of the parameters of
%the narrow component, because $\epsilon_{e,-1,n}$ cannot be very
%different from 3.3 and, if we decreased $\epsilon_{B,-2,n}$ we would
%have to increase $E_{52,0,n}$ by the same amount (see
%Eq.~\ref{a500}), and the second member of Eq.~\ref{a1200} would not
%change. This argument is important, because we wish to avoid
%unphysically high values of kinetic energy of the wide component at
%late times.

By rearranging Eq. \ref{b} and \ref{a100} we derive

\begin{equation}
\epsilon_{e,-1,w}^{-4} \epsilon_{B,-2,w}^{-0.5} > 3\times 10^{3}
\end{equation}
which, with the value of $\epsilon_{B,-2,w}$ chosen above,  implies
$\epsilon_{e,-1,w}<0.1$.

Let us assume the following series of parameters: $n = 3 \times
10^{-3}$, $\epsilon_{e,-1,n}=3.3$, $\epsilon_{B,-2,w}=10$,
$\epsilon_{e,-1,w}=0.04$ and $E_{52,0,w}=600$. As we will explain in
the following, larger values of $E_{52,0,w}$ are implausible, since
they translate into  unphysically high values of kinetic energy of
the wide component at late times. Moreover, since our model requires
a substantial difference in the beaming angles of the wide and
narrow components, then the difference in the respective kinetic
energies must not be too large. We then assume $E_{52,0,n} \approx
0.20 E_{52,0,w} = 120$, and $\epsilon_{B,-2,n}=0.1$ (the latter to
satisfy Eq.\ref{a1000}). This set of parameter values satisfies all
the required inequalities. We note that the narrow component has
``standard'' values of the two $\epsilon$'s (Panaitescu \& Kumar
2001a, 2001b). The wide component, instead, should have an
inefficient conversion of shock energy into electron energy and a
very efficient conversion of shock energy into magnetic field.
%Alternatively, the wide outflow might be endowed with strong
%magnetic field by the central engine; see Zhang and Meszaros (2002).
Furthermore, the wide component should carry a high amount of
energy, since $E_{K,w}$ is already as high as  $6 \times
10^{54}$~ergs 300 seconds after the trigger, and it increases, in
our model, as $t^{\sim0.5}$. As mentioned above, $E_{K, w}$ should
not be much higher than this value. For example, if the initial
value of the wide component kinetic energy is $E_{0,w} =
10^{56}$~ergs, this quantity would become as large as $E\sim 3
\times 10^{57}$~ergs 4 days after the trigger.  This very high value
would likely pose a energy budget problem for the central engine of
the GRB. If GRB optical lightcurve undergoes a jet break several
days after the trigger (Frail et al. 2001), for these very high
values of kinetic energy the beaming correction would be $\sim
10^{-4}$ (see Eq. \ref{9}). If, instead, $E_{0,w} = 6 \times
10^{54}$, the kinetic energy of the wide outflow would approach
$10^{56}$~ergs 1 day after the trigger, and $2\times10^{56}$~ergs 4
days after the trigger. If corrected for the beaming factors seen
above, the energy of the wide component would be of order of
$10^{52}$~ergs which, although high, is still acceptable according
to GRB theoretical models. The large majority of ejecta kinetic
energy is carried by the wide outflow. Since, in the prompt emission
phase, the GRBs emit isotropically around $10^{53}$~ergs in gamma
ray, values of efficiency $\eta$ as low as a fraction of percent
should be assumed (Zhang \& Meszaros 2004), at least for the wide
outflow.\newline
 A possible limit of scenario A' is that, even by assuming a value
of the circumburst medium density as low as $\sim 3 \times 10^{-3}$,
it requires a certain degree of fine tuning between the parameters.
The inequalities required by this scenario can be solved also for
slightly larger values, i.e. $n\sim 10^{-2}$, but the allowed region
in the parameters space becomes smaller and even finer tuning is
needed.

\subsection{Scenario A''}

 We will now explore a variant of scenario A', which is obtained
by placing the synchrotron peak frequency of the narrow component
below the optical band. This condition must now hold since 300~s.
The new dataset of inequalities reads

\begin{equation}\label{a16}
    f_{O,w} > 2 f_{O,n}
\end{equation}

\begin{equation}\label{a17}
   f_{X,w} < \frac{1}{2} f_{X,n} \qquad {\rm at \, 300 \, s \, after \, the
\, trigger.}
\end{equation}

\begin{equation}\label{a18}
  \nu_{m,w} < \nu_O  \qquad
\end{equation}

\begin{equation}\label{a19}
\nu_{O} < \nu_{c,w} \qquad {\rm at \, 0.1 \, d \, after \, the \,
trigger.}
\end{equation}

\begin{equation}\label{a20}
  \nu_{x} > \nu_{c,w}  \qquad {\rm at \, 300 \, s \, after \, the
\, trigger}
\end{equation}

%\begin{equation}\label{a21}
%  \nu_{x} > \nu_{m,n}  \qquad {\rm at \, 300 \, s \, after \, the
%\, trigger}
%\end{equation}

\begin{equation}\label{a22}
  \nu_{m,n} < \nu_{O}  \qquad {\rm at \, 300 \, s \, after \, the
\, trigger}
\end{equation}

\begin{equation}\label{a23}
  \nu_{x} < \nu_{c,n}  \qquad {\rm at \, 0.1 \, d \, after \, the
\, trigger \, .}
\end{equation}

 We note that the inequality $f_{O,w} > 2 f_{O,n}$ now has no requirement on
time, since it deals with fluxes in the same spectral regime.
However, its expression will have to change from the previous
scenario. Eq.~\ref{a22} also changes. We have

\begin{equation}\label{16000}
\epsilon_{B,-2,w} ^{0.85} E_{52,0,w} ^{1.35} \epsilon_{e,-1,w}^{1.4}
> 2 \epsilon_{B,-2,n} ^{0.85} E_{52,0,n} ^{1.35}
\epsilon_{e,-1,n}^{1.4}\, ,
\end{equation}

\begin{equation}\label{17000}
\epsilon_{B,-2,w} ^{0.1} \epsilon_{e,-1,w} ^{1.4} E_{52,0,w} ^{1.1}
< 2.85 \epsilon_{B,-2,n} ^{0.85} \epsilon_{e,-1,n} ^{1.4}
E_{52,0,n}^{1.35} n^{1/2}
\end{equation}

\begin{equation}\label{a18000}
\epsilon_{B,-2,w} ^{1/2} \epsilon_{e,-1,w} ^{2} E_{52,0,w} ^{1/2} <
0.22
\end{equation}

\begin{equation}\label{a19000}
 \epsilon_{B,-2,w}^{-3/2} E_{52,0,w} ^{-1/2} n^{-1} > 0.22
\end{equation}

\begin{equation}\label{a20000}
\epsilon_{B,-2,w} ^{-3/2} E_{52,0,w} ^{-1/2} n^{-1} < 32.5
\end{equation}

%\begin{equation}\label{a21000}
%\epsilon_{B,-2,n} ^{1/2} \epsilon_{e,-1,n} ^{2} E_{52,0,n} ^{1/2} <
%4\times 10^{2}
%\end{equation}

\begin{equation}\label{a22000}
\epsilon_{B,-2,n} ^{1/2} \epsilon_{e,-1,n} ^{2} E_{52,0,n} ^{1/2}
<0.22
\end{equation}

\begin{equation}\label{a23000}
\epsilon_{B,-2,n}^{-3/2} E_{52,0,n} ^{-1/2} n^{-1} > 4 \times 10^{2}
\, .
\end{equation}

 In this scenario, $\epsilon_{e,-1,n}$ should not be so high as in
other cases.
% The following values satisfy all the inequalities:
%$E_{52,0,n}=100$, $\epsilon_{e,-2,n}=0.25$,
%$\epsilon_{B,-2,n}=2\times10^{-3}$, $E_{52,0,w}=100$,
%$\epsilon_{e,-1,w}=0.25$, $\epsilon_{B,-2,w}=2$, $n=0.75$.
Equations \ref{a800} and \ref{a900} still apply. In this scenario,
it is possible to have values of $E_{52,0,w}$ much lower than in the
previous scenarios and well below $E_{52,0,n}$. In fact, these
values meet all the posed conditions: $E_{52,0,n}=100$,
$\epsilon_{e,-2,n}=0.25$, $\epsilon_{B,-2,n}=2\times10^{-3}$,
$E_{52,0,w}=3$, $\epsilon_{e,-1,w}=0.25$, $\epsilon_{B,-2,w}=2$,
$n=0.5$.  This fact has important consequences. In our modelling, it
is intrinsically assumed that we have emissions from spherical
portions of two outflows, and the emitting surface of the narrow
outflow is much less than the surface of the wide outflow. This
approximation can hold if the beaming angles of the two outflows are
different enough. If $\theta_{W} \simeq \theta_{N}$ the emitting
surface of the wide outflow would be better approximated by a ring
rather than a portion of spherical surface. This configuration would
lead to a behaviour of the optical emission which is different from
that described in our scenario. Now, in the previous scenarios, any
break in the optical should be much later than the chromatic break
in the X-ray, otherwise, from Eq.~\ref{9}, we would have indeed
drawn that $\theta_{W} \simeq \theta_{N}$. This stems from the fact
that in all previous scenarios $E_{52,0,w}$ is much higher than
$E_{52,0,n}$. However, in Scenario A'', it is $E_{52,0,w} \simeq
0.03 E_{52,0,n}$. Therefore, $\theta_{W} > \theta_{N}$ even if any
jet break in the optical occurs slightly after the jet break in the
X-ray. This case can be applied, for example, to GRB060605. Thus, we
conclude that Scenario A'' fits better the cases of GRBs that show
optical breaks only slightly later than the break in the X-ray. \\
Note, though, that scenario A'' can be solved even with high values
of the kinetic energies. The following choice of parameters satisfy
the conditions: $E_{52,0,n}=3\times10^3$, $\epsilon_{e,-1,n} = 0.1$,
$\epsilon_{B,-2,n}=2\times10^{-3}$, $E_{52,0,w}=200$, $\epsilon_{e,-1,w}=0.1$,
$\epsilon_{B,-2,w}$, $n=0.5$.\\
 A possible advantage of this scenario is that it does not
necessarily require high values of kinetic energy of the ejecta, so
it can be applied to dim bursts and/or bursts with higher efficiency
$\eta$ with respect to other models.\\

 As a potential drawback, in Scenario A'' fine tuning is not removed,
because a few inequalities are satisfied within factors of 1.5-2.

%Therefore, this scenario can accommodate the cases of GRBs that show
%optical breaks slightly later than the break in the X-ray. Since
%$E_{52,0,w} \sim 0.01 E_{52,0,n}$, we have $\theta_{W} > 2.5
%\theta_{N}$, even if $t_O = 3 t_{X,2}$ (see Eq.\ref{9}). If
%$\theta_{W}$ became comparable with $\theta_N$, the surface of the
%wide outflow should be more correctly approximated to a annulus
%rather than a portion of spherical surface, and this would produce
%different behaviours from those we have predicted so far.
%coming from the wide component should be treated as an annulus
%  Fine

\subsection{Scenario B}

In this scenario, the conditions to be fulfilled are:

\begin{equation}\label{b1}
    f_{O,w} > 2 f_{O,n}
\end{equation}

\begin{equation}\label{b2}
   f_{X,w} < \frac{1}{2} f_{X,n}
\end{equation}

\begin{equation}\label{b4}
  \nu_{x} > \nu_{c,w}  \qquad {\rm at \, 300 \, s \, after \, the
\, trigger}
\end{equation}

\begin{equation}\label{b7}
  \nu_{m,w} < \nu_O  \qquad {\rm at \, 300 \, s \, after \, the
\, trigger.}
\end{equation}

\begin{equation}\label{b7a}
  \nu_{c,w} > \nu_O  \qquad {\rm at \, 0.1 \, d \, after \, the
\, trigger, }
\end{equation}

\begin{equation}\label{b3}
  \nu_{x} > \nu_{c,n}  \qquad {\rm at \, 300 \, s \, after \, the
\, trigger}
\end{equation}

\begin{equation}\label{b5}
  \nu_{x} > \nu_{m,n}  \qquad {\rm at \, 300 \, s \, after \, the
\, trigger}
\end{equation}

\begin{equation}\label{b6}
  \nu_{m,n} > \nu_{O}  \qquad {\rm at \, 0.1 \, d \, after \, the
\, trigger}
\end{equation}

which  now give

\begin{equation}\label{b8}
\epsilon_{B,-2,w} ^{0.85} E_{52,0,w} ^{1.35} \epsilon_{e,-1,w}^{1.4}
> 29.5 \epsilon_{B,-2,n} ^{1/3} E_{52,0,n} ^{5/6}
\epsilon_{e,-1,n}^{-\frac{2}{3}}\, ,
\end{equation}

\begin{equation}\label{b9}
\epsilon_{B,-2,w} ^{0.1} E_{52,0,w} ^{1.1} \epsilon_{e,-1,w}^{1.4} <
\frac{1}{2} \epsilon_{B,-2,n} ^{0.1} E_{52,0,n} ^{1.1}
\epsilon_{e,-1,n}^{1.4}\, ,
\end{equation}

\begin{equation}\label{b11}
 \epsilon_{B,-2,w} ^{-3/2} E_{52,0,w} ^{-1/2} n^{-1} < 32.5 \, ,
\end{equation}

\begin{equation}\label{b14}
 \epsilon_{B,-2,w} ^{1/2}  \epsilon_{e,-1,w} ^{2}
E_{52,0,w} ^{1/2} < 0.2  \, ,
\end{equation}

\begin{equation}\label{eqa}
\epsilon_{B,-2,w}^{-3/2} E_{52,0,w} ^{-1/2} n^{-1} > 0.22 \, .
\end{equation}

\begin{equation}\label{b10}
 \epsilon_{B,-2,n} ^{-3/2} E_{52,0,n} ^{-1/2} n^{-1} < 32.5\, ,
\end{equation}

\begin{equation}\label{b12}
 \epsilon_{B,-2,n} ^{1/2}  \epsilon_{e,-1,n} ^{2}
E_{52,0,n} ^{1/2}  < 4 \times 10^2 \, , \end{equation}

\begin{equation}\label{b13}
 \epsilon_{B,-2,n} ^{1/2}  \epsilon_{e,-1,n} ^{2}
E_{52,0,n} ^{1/2} >14.7 \, ,
\end{equation}

By comparing Eq.~\ref{b8} and ~\ref{b9}, we can immediately infer
that $\epsilon_{e,-1,n}$ should be as high as possible, in order for
these two equations to be fulfilled more easily. Furthermore, a high
value of $\epsilon_{e,-1,n}$ makes easier to have the synchrotron
frequency of the narrow component higher than the optical band (Eq.
~\ref{b13}) even at late times. Therefore, we assume again  the
equipartition value of $\epsilon_{e,-1,n} = 3.3$, as in the previous
scenarios.

It is easy to check that Eq.~\ref{b13} gives limits less stringent
than Eqs.~\ref{b10},\ref{b12}. By combining the latter two, we can
see that they are satisfied by all plausible values of the energy of
the narrow component (since they only imply $E_{52,0,n} < \times 7.3
\times 10^{5} n $), while for the other parameters they give
\begin{equation}\label{b15}
\epsilon_{B,-2,n}^{-1/2} < 3 E_{52,0,n}^{1/6} n^{1/3} \, ,
\end{equation}
\begin{equation}\label{b16}
\epsilon_{B,-2,n}^{-1} n^{-1}< 1.2 \times 10^3 \, .
\end{equation}
 Therefore, as far as the narrow component is concerned, scenario B
requires that $n$ and $\epsilon_{B,-2,n}$ are not simultaneously
very small. For instance, for $E_{52,0,n} \sim 10$, it must be
$\epsilon_{B,-2,n} > 5 \times 10^{-2} n^{-2/3}$ and for value of
values of $\epsilon_{B,-2,n}$ the corresponding limit on $n$ must be
computed accounting for Eq.~\ref{b16} as well.

 Stringent limits on the $\epsilon_{B,-2,w}$ can be obtained by
considering the conditions of the wide component. By using
Eq.~\ref{b11} and ~\ref{b14}, we have
\begin{equation}\label{e1}
E_{52,0,w} < 0.35 n \epsilon_{e,-1,w}^{-6} \, .
\end{equation}
%As we can see, a small $n$ implies tiny values of the electron
%energy. If  $n\sim 0.1$ and we assume an energy of $E_{52,0,w} \sim
%2 \times 10^2$, the condition above is compatible with an upper
%limit of $\epsilon_{e,-1,w} < 0.3$.

 As we can see, $\epsilon_{e,-1,w}$ should be quite small, in order
to permit values of kinetic energy of the wide outflow that are
comparable with those observed in a few luminous GRBs, of the order
$\approx 10^{54}$~ergs (Frail et al. 2001). For instance, if $n=0.1$
and $E_{52,0,w}=250$, then $\epsilon_{e,-1,w} < 0.2$. In the
following we will assume $\epsilon_{e,-1,w} = 0.06$. Finally, an
upper limit on $\epsilon_{B,-2,w}$ can be then obtained from
Eq.~\ref{b14}, $\epsilon_{B,-2,w} < 0.05 \epsilon_{e,-1,w}^{-4}
E_{52,0,w}^{-1}$ which, with our choice of the parameter values,
requires $\epsilon_{B,-2,w}< 40$.

 It can be easily shown that, by  using $E_{52,0,n}=30$, $n=0.005$,
$\epsilon_{B,-2,n} = 1.5$, $\epsilon_{e,-1,n} = 3.3$,
$E_{52,0,w}=300$, $\epsilon_{B,-2,w} = 10$, $\epsilon_{e,-1,w}=0.06$
all the required conditions are satisfied. We notice that this
scenario again requires a large degree of fine tuning between the
parameters. It also requires a high value of kinetic energy of the
wide component, almost as high as in Scenario A'. It therefore
requires that the efficiency of conversion of this kinetic energy
into $\gamma$-rays is as low as in A'. Furthermore, in Scenario B,
the relative ratio of the two component isotropic energies is
$E_{52,n} / E_{52,w} \sim 10\%$, i.e. lower than in Scenario A'.
Such a large difference in the two energies might cause the beaming
angles of the two outflow not to differ considerably, unless a jet
break in the optical occurs much later than the break in the X-ray
(see Eq.~\ref{9}).

\subsection{Scenario B'}

We will now explore a variant of the previous case, in which the
cooling frequency of the wide component lie above the X-ray band. We
therefore reverse condition ~\ref{b4}. Notice that the time when
this condition has to hold changes as well; it can be shown that, in
this scenario, if it holds at 0.1 days then it also holds since the
beginning. Note also that expression \ref{b9}, that relates to
condition \ref{b2}, has to be changed as well.\\
Overall, the required conditions now read:

\begin{equation}\label{b17}
    f_{O,w} > 2 f_{O,n}
\end{equation}

\begin{equation}\label{b18}
   f_{X,w} < \frac{1}{2} f_{X,n}
\end{equation}

\begin{equation}\label{b19}
  \nu_{x} > \nu_{c,n}  \qquad {\rm at \, 300 \, s \, after \, the
\, trigger}
\end{equation}

\begin{equation}\label{b20}
  \nu_{x} < \nu_{c,w}  \qquad {\rm at \, 0.1 \, d \, after \, the
\, trigger}
\end{equation}

\begin{equation}\label{b21}
  \nu_{x} > \nu_{m,n}  \qquad {\rm at \, 300 \, s \, after \, the
\, trigger}
\end{equation}

\begin{equation}\label{b22}
  \nu_{m,n} > \nu_{O}  \qquad {\rm at \, 0.1 \, d \, after \, the
\, trigger}
\end{equation}

\begin{equation}\label{b23}
  \nu_{m,w} < \nu_O  \qquad {\rm at \, 300 \, s \, after \, the
\, trigger \, , }
\end{equation}
which translate into:

\begin{equation}\label{b24}
\epsilon_{B,-2,w} ^{0.85} E_{52,0,w} ^{1.35} \epsilon_{e,-1,w}^{1.4}
> 29.5 \epsilon_{B,-2,n} ^{1/3} E_{52,0,n} ^{5/6}
\epsilon_{e,-1,n}^{-2/3}\, ,
\end{equation}

\begin{equation}\label{b25}
\epsilon_{B,-2,w} ^{0.85} E_{52,0,w} ^{1.35} \epsilon_{e,-1,w}^{1.4}
n^{1/2} < 2.5 \times 10^{-2} \epsilon_{B,-2,n} ^{0.1} E_{52,0,n}
^{1.1} \epsilon_{e,-1,n}^{1.4}\, ,
\end{equation}

\begin{equation}\label{b26}
 \epsilon_{B,-2,n} ^{-3/2} E_{52,0,n} ^{-1/2} n^{-1} < 32.5\, ,
\end{equation}

\begin{equation}\label{b27}
 \epsilon_{B,-2,w} ^{-3/2} E_{52,0,w} ^{-1/2} n^{-1} > 4 \times 10^2 \, ,
\end{equation}

\begin{equation}\label{b28}
 \epsilon_{B,-2,n} ^{1/2}  \epsilon_{e,-1,n} ^{2}
E_{52,0,n} ^{1/2}  < 4 \times 10^2 \, ,
\end{equation}

\begin{equation}\label{b29}
 \epsilon_{B,-2,n} ^{1/2}  \epsilon_{e,-1,n} ^{2}
E_{52,0,n} ^{1/2} >14.7 \, ,
\end{equation}

\begin{equation}\label{b30}
 \epsilon_{B,-2,w} ^{1/2}  \epsilon_{e,-1,w} ^{2}
E_{52,0,w} ^{1/2} < 0.22  \, .
\end{equation}

As it can be easily seen, the simultaneous validity of both
inequalities \ref{b24} and ~\ref{b25} (which have similar left
members apart from the factor $n^{1/2}$) crucially depends on the
value of $\epsilon_{e,-1,n}$, that must be relatively large.
Therefore in the following we assume again $\epsilon_{e,-1,n}=3.3$.
Once $\epsilon_{e,-1,n}$ has been assigned, Eq.~\ref{b24} and
~\ref{b25} are more easily satisfied for relatively low values of
the density and of $\epsilon_{B,-2,n}$ and for  relatively high
values of the kinetic energy of the narrow component. Besides, since
$\epsilon_{e,-1,n}=3.3$, relations \ref{b15} and \ref{b16},
involving the narrow component only, still apply. Based on that, we
assume the following set of parameters for the narrow component:
$E_{52,0,n}=4000$, $\epsilon_{e,-1,n}=3.3$, $\epsilon_{B,-2,n}=0.2$
and a density $n=10^{-2}$.  With these choices, some of the
Eqs.~\ref{b24}-\ref{b29} are trivially satisfied, while the others
give

\begin{equation}\label{b24''}
\epsilon_{B,-2,w} ^{0.85} E_{52,0,w} ^{1.35} \epsilon_{e,-1,w}^{1.4}
> 7.5\times 10^{3} \, ,
\end{equation}

\begin{equation}\label{b25''}
\epsilon_{B,-2,w} ^{0.85} E_{52,0,w} ^{1.35} \epsilon_{e,-1,w}^{1.4}
n^{1/2} < 1.05 \times 10^3 \, ,
\end{equation}

\begin{equation}\label{b27''}
\epsilon_{B,-2,w} ^{-3/2} E_{52,0,w} ^{-1/2} > 4  \, .
\end{equation}

From Eq.~\ref{b30} we can isolate an expression for
$\epsilon_{e,-1,w}$ which,  substituted  into Eq.\ref{b24''}, gives

\begin{equation}\label{b33}
\epsilon_{B,-2,w}^{0.5} E_{52,0,w} > 2.3 \times 10^4 \, .
\end{equation}

From this last equation we can immediately infer a lower limit on
the value of $E_{52,0,w}$. Since the highest theoretical value of
$\epsilon_{B,-2,w}$ is 33, achieved at equipartition, the minimum
value of $E_{52,0,w}=4\times10^3$, which is admittedly very high. By
using this value of $E_{52,0,w}$ in Eq.~\ref{b27''}, we derive an
upper limit on $\epsilon_{B,-2,w} < 6 \times 10^{-2}$. Also, we can
obtain an upper limit on $\epsilon_{e,-1,w}$ by solving
Eq.~\ref{b33} for $\epsilon_{B,-2,w} $ and substituting the
resulting expression into Eq.~\ref{b30}. We obtain:

\begin{equation}\label{b34}
\epsilon_{e,-1,w}^2 E_{52,0,w}^{-1/2} < 9.6 \times 10 ^{-6} \, .
\end{equation}

By using the upper limit on $E_{52,0,w}$ quoted above, this last equation
gives $\epsilon_{e,-1,w} < 2.5 \times 10^{-2}$. It is easy to verify
that, for these values of  the parameters of the wide outflow,
Eq.~\ref{b24''} cannot be satisfied, unless $E_{52,0,w}$ is
unphysically large, $\sim 10^{5}$. Scenario B' therefore cannot be
assumed in our model.

\subsection{Scenario B''}

We will now explore a variant of scenario B, in which the
synchrotron peak frequencies of both components are below the
optical band, and the cooling frequencies are between the optical
and the X-ray band. Overall, the required conditions now read:

\begin{equation}\label{b35}
    f_{O,w} > 2 f_{O,n}
\end{equation}

\begin{equation}\label{b36}
   f_{X,w} < \frac{1}{2} f_{X,n}
\end{equation}

\begin{equation}\label{b37}
  \nu_{m,w} < \nu_O  \qquad {\rm at \, 300 \, s \, after \, the
\, trigger \, , }
\end{equation}

\begin{equation}\label{b38}
  \nu_{x} > \nu_{c,w}  \qquad {\rm at \, 300 \, s \, after \, the
\, trigger}
\end{equation}

\begin{equation}\label{b39}
  \nu_O < \nu_{c,w}  \qquad {\rm at \, 0.1 \, d \, after \, the
\, trigger}
\end{equation}

\begin{equation}\label{b40}
  \nu_{O} > \nu_{m,n}  \qquad {\rm at \, 300 \, s \, after \, the
\, trigger}
\end{equation}

\begin{equation}\label{b41}
  \nu_{c,n} < \nu_{X}  \qquad {\rm at \, 300 \, s \, after \, the
\, trigger}
\end{equation}

\begin{equation}\label{b42}
  \nu_{c,n} > \nu_O  \qquad {\rm at \, 0.1 \, d \, after \, the
\, trigger \, , }
\end{equation}

which translate into:

\begin{equation}\label{b43}
\epsilon_{B,-2,w} ^{0.85} E_{52,0,w} ^{1.35} \epsilon_{e,-1,w}^{1.4}
> 2 \epsilon_{B,-2,n} ^{0.85} E_{52,0,n} ^{1.35}
\epsilon_{e,-1,n}^{1.4}\, ,
    \end{equation}

\begin{equation}\label{b44}
\epsilon_{B,-2,w} ^{0.1} E_{52,0,w} ^{1.1} \epsilon_{e,-1,w}^{1.4}
> \frac{1}{2} \epsilon_{B,-2,n} ^{0.1} E_{52,0,n} ^{1.1}
\epsilon_{e,-1,n}^{1.4}\, ,
\end{equation}

\begin{equation}\label{b45}
\epsilon_{B,-2,w}^{1/2} E_{52,0,w}^{1/2} \epsilon_{e,-1,w}^2 < 0.22
\end{equation}

\begin{equation}\label{b46}
 \epsilon^{-3/2} E_{52,0,w}^{-1/2} n^{-1} > 0.22
\end{equation}

\begin{equation}\label{b47}
\epsilon_{B,-2,w}^{-3/2} E_{52,0,w}^{-1/2} n^{-1} < 32.5
\end{equation}

\begin{equation}\label{b48}
\epsilon_{B,-2,n}^{1/2} E_{52,0,n}^{1/2} \epsilon_{e,-1,n} ^{2} <0.22
\end{equation}

\begin{equation}\label{b49}
\epsilon_{B,-2,n}^{-3/2} E_{52,0,n} ^{-1/2} n^{-1/2} > 0.22
\end{equation}

\begin{equation}\label{b50}
\epsilon_{B,-2,n} ^{-3/2} E_{52,0,n}^{-1/2} n^{-1} < 32.5
\end{equation}

 Placing a different condition $f_{O,w} > 2 f_{O,n}$ and because
$\nu_{m,n}$ is below the optical band since the very beginning, it
is no longer necessary to assume a high value for
$\epsilon_{e,-1,n}$ to simultaneously satisfy the first two
conditions. Instead, by combining Eq.\ref{b48} and \ref{b49}, we
obtain

\begin{equation}\label{b51}
E_{52,0,n} < 3.5 \epsilon_{e,-1,n}^{-6} n
\end{equation}

From this inequality, we derive that $\epsilon_{e,-1,n}$ should be
small, to allow high values of kinetic energy of the narrow outflow.
As for the wide outflow, condition Eq.\ref{e1} still applies.\\

For the following values of parameters, all the relevant
inequalities of scenario B'' are satisfied: $E_{52,0,w}=0.25$, $
\epsilon_{e,-1,w}=0.25$, $\epsilon_{B,-2,w}=10$, $E_{52,0,n}=0.5$,
$\epsilon_{e,-1,n}=0.25$, $\epsilon_{B,-2,n} = 0.3$, $n=0.75$.
Scenario B'' can be solved for higher values of kinetic energies as
well: $E_{52,0,w}=20$, $ \epsilon_{e,-1,w}=0.1$,
$\epsilon_{B,-2,w}=5$, $E_{52,0,n}=90$, $\epsilon_{e,-1,n}=0.1$,
$\epsilon_{B,-2,n} = 0.15$, $n=0.75$ satisfy all conditions.\\

Scenario B'' is similar to Scenario A'', in the sense that it can be
resolved for high and low values of the kinetic energies, and even
in this case, $E_{52,0,w} < E_{52,0,n}$. Likewise, Scenario B'' does
not solve the problem of fine tuning, and the ratio of $E_{52,0,w} /
E_{52,0,n}$ is is much higher than in A''. This scenario cannot thus
be employed for cases in which a jet break occurs in the optical
slightly after the jet break in the X-ray. Furthermore, it still
presents the problem of fine tuning.

\subsection{Summary}
\label{sum}

In summary, we have shown that there are at least two scenarios of
``A'' kind and two of ``B'' kind that are satisfied for
non-unreasonable values of the parameters. A drawback is that in all
cases we require a large degree of fine tuning, since the allowed
region in the parameter space is small. Since the bursts with
chromatic breaks may not be rare (Liang et al. 2008), fine tuning
can represent a problem for our model.

% This fact,
%however, could explain the rarity of GRBs with chromatic breaks
%(Oates et al. in preparation, Melandri et al. 2008).
% (\textsl{thought: but GRBs with chromatic breaks are rare,
%aren't they?})
 We would like now to address the point of the reliability of our
model at late times, i.e. after the break observed at $0.1$ days
after the trigger. Within our model, this break is interpreted as a
jet-break. This implies that, from this time onwards, the flux of
the narrow component is expected to decrease considerably faster
than before, while the flux due to the wide outflow does not change
its decay slope. Therefore, it is important to check that the flux
in the X-ray due to the narrow component remains above that due to
the wide component even at late times. Would this condition not be
satisfied we should observe  a flattening of the X-ray lightcurve,
as $f_{X,w}$ becomes comparable to $f_{X,n}$ at some time after the
end of the plateau phase; this is clearly not observed in our GRB
lightcurves.

Now, for $p=2.4$, the X-ray flux of the narrow component decreases
with time as $f_{X,n} \propto t^{-1.5}$ in Scenario A' and A'', and
$f_{X,n} \propto t^{-1.67}$ in B  and B''respectively. $f_{X,w}$
always decays as $\sim t^{-0.75}$. Therefore, the ratio $f_{X,n} /
f_{X,w} $ will decrease as $t^{-0.75}$ in scenario A' and A'' and as
$t^{-0.9}$ in scenario B and B''. With our suggested choice of
parameters, condition $f_{X,n}> f_{X,w}$ is satisfied (by a factor
of $\sim10$) in scenario A' at 0.1~d after the trigger, suggesting
that a flattening of the X-ray lightcurve will not be seen before
2-2.5 days after the trigger, when lightcurves are usually poorly
sampled. In scenario A'', $f_{X,n}> f_{X,w}$ by a factor $\sim2.5$
only at 0.1~d after the trigger; therefore, in this case, the X-ray
flux of the wide component becomes comparable with that produced by
the narrow outflow as early as $\sim0.3$~d after the trigger and the
X-ray decay slope should become similar to that in the optical,
unless an early jet break occurs in the wide outflow as well.

 In the case of Scenario B, $f_{X,n} > f_{X,w}$ is satisfied by a
factor of $\sim20$ 0.1~d after the trigger. One should thus expect a
flattening as late as in Scenario A''. Finally, in Scenario B'',
$f_{X,n} \sim 2.5 f_{X,w}$ for our choice of parameters, therefore
the same restrictions of Scenario A'' apply in this case, too.

%Thus, the expected time for any flattening in the X-ray
%curve is only slightly beyond that of Scenario B.

% This indicates that, at
%about $\sim1$d after the trigger, the X-ray flux of the wide
%component becomes comparable with that produced by the narrow
%outflow, and the X-ray decay slope should become similar to that in
%the X-ray, unless a jet break occurs in the wide outflow as well.

 In drawing our scenario, we restricted ourselves to the simplest
case of side-spreading jets and a constant density medium, with the
addition of energy injection (Panaitescu et al. 2006b) parameterized
as $L\propto t^{0.5}$, and fixed $p=2.4$. We also assumed a simple
hierarchy between the relevant frequencies. In this simplified case,
we have shown that our model successfully explains the
characteristics of all bursts in our sample, with the only
difference that in some cases we need to assume $\nu_X>\nu_c$, and
in some others the reversed inequality.
% Both cases, however, would
%imply GRBs with very high energy budget that explode in relatively
%small values of the particle density GRBs (see, e.g., Panaitescu \&
%Kumar 2001a).
In many cases, the fraction of energy of the narrow outflow given to
the emitting electrons has to be close to the maximum value allowed
for adiabatic expansion (Freedman \& Waxman 2001, Yost et al. 2003).
In the case of GRBs without well sampled optical emission, we have
deemed not to assume the Scenarios A'' and B'', which would require
the presence of a flattening of the X-ray lightcurve only a fraction
of day after the trigger, which is not observed.

% Cases requiring $\nu_X < \nu_c$ for the whole
%duration of the observation, in turn, would imply relatively small
%values of the particle density
% and/or the
%magnetic energy,
%while cases with $\nu_X > \nu_c$ should involve GRBs with very high
%energy budget (see, e.g. Panaitescu \& Kumar 2001a). In all cases,
%the fraction of energy of the narrow outflow given to the emitting
%electrons has to be close to the maximum value allowed for adiabatic
%expansion (Freedman \& Waxman 2001, Yost et al. 2003).
% and no break observed in the optical till several days
%after the trigger. .
It is worth mentioning that we have also explored Scenarios A', B,
A'' and B'' in a wind scenario. We adopted the same frequency
hierarchies of these two cases, but we replaced the set of equations
\label{eqnset1} with the set that describes the evolution of the
characteristics frequencies and peak flux in a wind environment.
These formulae were taken from Yost et al. 2003. We found that, even
in the case of circumburst medium environment, these four scenarios
basically reproduce the observed behaviours, but fine tuning is not
removed.
Of course, it is possible to apply more complicated
scenarios. For example, we may choose values of the parameters $q$
and $p$ which are different from those we have adopted in this
paper, to reflect intrinsic differences among the various bursts.
Changes of $p$ and $q$ from the values we have taken would result in
a modification of both the exponents and the coefficients of the
mathematical expressions we have used so far. As a consequence, some
scenarios might not be viable anymore, or others could become
applicable.
%  may reveal
%further intrinsic differences between the various bursts.

 We notice that our model can easily explain one of the most
striking characteristics of the GRBs studied by \textit{Swift}, i.e.
the lack of evident jet breaks in the X-ray lightcurves (Burrows \&
Racusin 2007). In our scenario jet breaks are actually observed, but
they are not so steep as we would expect from the traditional
closure relationships (Sari, Piran \& Helpern 1999) due to the
ongoing energy injection. Our model predicts that the steep decay
slopes, like those observed in the optical in pre-\textit{Swift}
GRBs at late times, are possible only once the energy injection has
terminated.
%The hypothesis of a ``standard reservoir'' of kinetic
%energy in GRBs (Frail et al. 2001), based on the observation of
%optical jet breaks, is not in contradiction with our model, if most
%of the energy is carried by the wide outflow. In fact, in our
%modelling, we have always found that most of the kinetic energy of
%the ejecta is carried by the wide outflow. We note, however, that in
%our model the value of the standard energy reservoir tends to be
%higher than in Frail et al., while beaming angles and efficiencies
%are lower. Furthermore, energy itself is a fine-tuned parameter.
We note that our model might, in principle, be extended to all GRBs
featuring the canonical lightcurve (Nousek et al. 2006), even those
without chromatic breaks. The implication would be that, in those
cases where optical and X-rays lightcurves show a simultaneous break
at the end of the slow decline phase, the emission in both bands
would arise from the same outflow. However, in our scenario the
break is not caused by the end of an energy injection phase, as
generally assumed when interpreting the canonical lightcurve, but by
a jet break. Once the energy injection has terminated, the decline
slope of optical and X-ray fluxes will assume the more typical
values of $\alpha \simeq 2$. Thus our model can also explain GRBs
which show achromatic breaks only. The values of the decay and
spectral slopes of the GRBs we have studied in this paper are not
uncommon, supporting the idea that our model could be applied in
several cases. Our interpretation can call for a deep revision of
GRB physics, such as the mechanism that produces the outflow and the
energetics involved in the process. We need to explain how the
central engine can either be active for several days, or produce a
long trail of shells that merge for such a long time. Besides, we
should find mechanisms that can commonly beam ejecta into cones,
which can have opening angles as narrow as $5\times10^{-3}$ rad.

\section{Conclusions}
\label{conc}

 In this paper, we have reanalysed the full sample of \textit{Swift}
GRBs whith chromatic breaks, originally discussed by Panaitescu et
al. (2006a). In addition, we have also studied GRB 060605, another
\textit{Swift} burst with good quality XRT and UVOT data and a
chromatic break in the XRT lightcurve.
 We have shown how our model, based on a prolonged energy injection into a
double component outflow and a jet break, is physically plausible
and can well explain the behaviour of the optical and X-ray emission
of GRB050319, 060505 and GRB050802 (see also Oates et al. 2007).
%While GRB060505 shows that the optical and X-ray emission should be
%emitted by physically different mechanisms, this burst cannot be
%interpreted in our model, at least with the simplest assumptions.
GRB050922c has been shown not to require a chromatic break. We note
that our model can also be applied to the other GRBs with claim of
chromatic breaks published in Panaitescu et al.~(2006a) and might,
in principle, be extended to all GRBs featuring the canonical
lightcurve (Nousek et al. 2006), even those without chromatic
breaks.
% we remind, though,
%that some degree of tuning of physical parameters values are needed.
We emphasize that it would have not been possible to derive our
conclusions if we had considered the X-ray data only, since
GRB050319, GRB060605 and GRB050802 exhibit a canonical X-ray
lightcurve. Instead, the combined optical and X-ray analysis has
shown that the component responsible for the optical is uncoupled
from the outflow that produces the X-ray emission. In our model, the
ejecta responsible for the X-ray emission are narrowly beamed, and
undergo an early jet break that explains the chromatic break seen in
the X-ray only. Our model of combined jet expansion and energy
injection may have deep consequences on our understanding of the
GRB, since it calls for a revision of the physics processes that
take place in these objects.
% Our model could be applied to all GRBs, including those displaying
%the ``usual'' achromatic breaks, and may thus have deep consequences
%on our understanding of the GRB physics.

\section{Acknowledgements}

 We thank B. Zhang for a comments and suggestions on this
manuscript.

\clearpage

\clearpage


\begin{thebibliography}{}


\bibitem[\protect\citeauthoryear{Band et al.}{1993}]{ban93}
Band D., 1993, \apj, 413, 281

\bibitem[\protect\citeauthoryear{Berger et al.}{2003}]{b03}
Berger E., Kulkarni S.R. \& Frail D.A., 2003, \apj, 590, 379

\bibitem[\protect\citeauthoryear{Bevington et al.}{1969}]{bev69}
Bevington, P.R., 1969, Data Reduction and Error Analysis for the
Physical Sciences, Ed. McGraw-Hill

\bibitem[\protect\citeauthoryear{Burrows et al.}{2004}]{bu04}
Burrows D., Proc. SPIE, 5165, 201

\bibitem[\protect\citeauthoryear{Burrows \& Racusin}{2007}]{bur07}
Burrows D.N., \& Racusin, J., 2007, in the proceeding of the ``Swift and
GRBs:
Unveiling the Relativistic Universe'' meeting,Venice, June 5-9, 2006,
published by "Il Nuovo Cimento", pag. 1273 (arXiv:astro-ph/0702633)

\bibitem[\protect\citeauthoryear{Chevalier \& Li}{2000}]{chl00}
Chevalier R.A. \& Li Z.Y., 2000, ApJ, 536, 195

\bibitem[\protect\citeauthoryear{Curran et al.}{2007}]{cur07}
Curran P.A., Van der Horst A.J, Wijers R.A.M.J. et al., 2007,
\mnras, 381, 65

\bibitem[\protect\citeauthoryear{Cusumano et al.}{2006}]{cus06}
Cusumano G. et al., 2006, \apj, 639, 316


\bibitem[\protect\citeauthoryear{De Pasquale et al.}{2006}]{depa05}
De Pasquale M., Piro L., Gendre B. et al., 2006, \aap, 455, 813

\bibitem[\protect\citeauthoryear{Evans et al.}{2007}]{eva07}
Evans P.A. et al., 2007, A\&A, 469, 379


\bibitem[\protect\citeauthoryear{Frail et al.}{2001}]{fr01}
Frail D.A., Kulkarni S.R., Sari R. et al., 2001, \apj, 562, L55


\bibitem[\protect\citeauthoryear{Gendre et al.}{2006}]{gen06} Gendre B. et
al., AIPC, 2006, 836, 558

\bibitem[\protect\citeauthoryear{Genet et al.}{2007}]{get07} Genet F. et
al., 2007, MNRAS, 381, 732

\bibitem[\protect\citeauthoryear{Gehrels et al.}{2004}]{geh06} Gehrels N.
et al., 2004 \apj, 611, 1005

\bibitem[\protect\citeauthoryear{Ghisellini et al.}{2007}]{ghi07}
Ghisellini G., Ghirlanda G., Nava L., Firmani C., 2007, \apjl, 658,
75

\bibitem[\protect\citeauthoryear{Godet et al.}{2008}]{god08}
Godet O. et al., 2008, SPIE in press (arXiv:0708.2988)

\bibitem[\protect\citeauthoryear{Golenetskii et al.}{2005}]{gol05}
Golenetskii S. et al., 2005, GCN 3179

\bibitem[\protect\citeauthoryear{Jacobsson et al.}{2006}]{jac06}
Jakobsson P., Levan A., Fynbo J.P.U. et al., 2006, \aap, 447, 897

\bibitem[\protect\citeauthoryear{Jaunsen et al.}{2001}]{jau02}
Jaunsen A.O. et al., 2001, \apj, 546, 127

\bibitem[{Kalberla et al.}{2005}]{kal05} Kalberla P. M. W.,
Burton W. B., Hartmann D., Arnal, E.M., Bajaja, E., Morras, R.,
Pöppel, W.G.L., 2005, A\&A, 440, 775

\bibitem[\protect\citeauthoryear{Kumar \& Panaitescu}{2000}]{kup00}
Kumar P. \& Panaitescu A., 2000, \apjl, 541, 51

\bibitem[\protect\citeauthoryear{Liang et al.}{2007}]{lia07}
Liang E.W. et al., 2007, \apj sub, astro-ph 07082942

\bibitem[\protect\citeauthoryear{Melandri et al.}{2008}]{mel08}
Melandri A. et al., 2008, \apj submitted, astro-ph 08040811

\bibitem[\protect\citeauthoryear{M\'{e}sz\'{a}ros \& Rees}{1997}]{mr97}
M\'{e}sz\'{a}ros P. \& Rees M.J., 1997, \apj, 476, 232

\bibitem[\protect\citeauthoryear{Micheal et al.}{1997}]{mic07}
Micheal R., Fordham J., \& Kawakami H., 1997, MNRAS, 292, 611

\bibitem[\protect\citeauthoryear{Moretti et al}{2006}]{mor06}
Moretti A. et al., 2006, AIPC, 836, 676

\bibitem[\protect\citeauthoryear{Morris et al.}{2007}]{mor07}
Morris D.C. et al., 2007, ApJ, 654, 413

\bibitem[\protect\citeauthoryear{Nousek et al.}{2006}]{nou06}
Nousek J. et al., 2006, \apj, 642, 389

\bibitem[\protect\citeauthoryear{O'Brien et al.}{2006}]{obr06}
O'Brien P. et al., 2006, \apj \/ submitted (astro-ph/0601125)

\bibitem[\protect\citeauthoryear{Oates et al.}{2007}]{oat07}
Oates. S. et al., 2007, MNRAS, 380, 270

\bibitem[\protect\citeauthoryear{Pagani et al.}{2006}]{pag06}
Pagani C. et al., 2005, AAS, 207, 7504

\bibitem[\protect\citeauthoryear{Panaitescu \& Kumar}{2001a}]{pak01a}
Panaitescu A. \& Kumar A., 2001, \apj 554, 667.

\bibitem[\protect\citeauthoryear{Panaitescu \& Kumar}{2001b}]{pak01b}
Panaitescu A. \& Kumar A., \apjl, 2001, 560 49

\bibitem[\protect\citeauthoryear{Panaitescu et al.}{2006a}]{pan06a}
Panaitescu A. et al., 2006a, MNRAS, 369, 2059 % CHROMATIC BREAK PAPER

\bibitem[\protect\citeauthoryear{Panaitescu et al.}{2006b}]{pan06b}
Panaitescu A. et al., 2006b, MNRAS, 366, 1357 % FORMULAE PAPER

\bibitem[\protect\citeauthoryear{Panaitescu}{2007a}]{pan07a}
Panaitescu A., 2007a, MNRAS, 380, 374

\bibitem[\protect\citeauthoryear{Panaitescu}{2007b}]{pan07b}
Panaitescu A., 2007b, MNRAS accepted (astro-ph/07081509)

\bibitem[\protect\citeauthoryear{Peng et al.}{2005}]{peng05}
Peng F., Königl A. et  Granot, J. 2005, \apj, 626, 966

\bibitem[\protect\citeauthoryear{Poole et al.}{2008}]{pol08}
Poole T.S. et al., 2008, MNRAS, 383, 627

\bibitem[\protect\citeauthoryear{Roming et al.}{2005}]{ro05}
Roming P. et al., 2005, SSRV, 120, 95.

\bibitem[\protect\citeauthoryear{Sato et al.}{2006}]{sat06}
Sato G. et al., 2006, GCN 5231

\bibitem[\protect\citeauthoryear{Sari et al.}{1998}]{spn98}
Sari R., Piran T. \& Narayan R., 1998, \apjl, 497, 17

\bibitem[\protect\citeauthoryear{Sari et al.}{1999}]{sph99}
Sari R., Piran T. \& Helpern J.P., 1999, \apjl, 519, 17

\bibitem[\protect\citeauthoryear{Schady et al.}{2007}]{sch07}
Schady P., Mason K.O., Page M.J., De Pasquale M. et al., 2007, MNRAS
377, 274

\bibitem[\protect\citeauthoryear{Stratta et al.}{2004}]{str04}
Stratta G., Fiore F., Antonelli L.A., Piro L., De Pasquale, M., 2004,
\apjl, 519, 17

\bibitem[\protect\citeauthoryear{Tagliaferri et al.}{2005}]{tal05}
Tagliaferri G., et al., 2005, Nature, 436, 985

\bibitem[\protect\citeauthoryear{Uhm \& Beloborodov}{2007}]{uhb07}
Uhm, Z.L. \& Beloborodov, A.M., 2007, \apjl, 665, 93

\bibitem[\protect\citeauthoryear{Yost et al.}{2003}]{yos03}
Yost. S. et al., 2003, \apj, 597, 459

\bibitem[\protect\citeauthoryear{Willingale et al.}{2007}]{wil07}
Willingale R., O'Brien P.T., Osborne J.P. et al., 2007, \apj, 662,
1093

\bibitem[\protect\citeauthoryear{Zhang \& Meszaros}{2004}]{zha04}
Zhang B., 2004, International Journal of Modern Physics A, Volume
19, Issue 15, pp. 2385

\bibitem[\protect\citeauthoryear{Zhang et al.}{2006}]{zh06}
Zhang B., Fan Y.Z., Dyks J. et al., 2006, \apj, 642, 354

\bibitem[\protect\citeauthoryear{Zhang}{2007}]{zh07}
Zhang B., 2007, Chinese Journal of Astronomy and Astrophysics,
Volume 7, Issue 1, pp. 1-50

\bibitem[\protect\citeauthoryear{Zhang et al.}{2007}]{zhal07}
Zhang B., Liang E., Page K.L. et al., 2007, \apj, 655, 989Z


\end{thebibliography}
\end{document}